\documentclass[reprint,prd,nofootinbib,amsmath,superscriptaddress,preprintnumbers,twocolumn,amssymb,aps,bibliography=totoc]{revtex4}[12pt]

\usepackage[usenames, dvipsnames, svgnames]{xcolor}
\usepackage[pdfview = FitH, pdfstartview = FitH, pdfremotestartview = FitH, pdfduplex = DuplexFlipLongEdge]{hyperref}
\hypersetup{
}

\usepackage{amsmath, amssymb, amsfonts}
\usepackage{graphicx}
\usepackage{epstopdf}
\usepackage{epsfig}
\usepackage[loose]{units}
\usepackage{latexsym}
\usepackage{verbatim}
\usepackage{dhucs}
\usepackage{dcolumn}
\usepackage{bm, bbm}	
\usepackage{notes2bib}
\usepackage{appendix}
\usepackage{slashed}
\usepackage{multirow}
\usepackage{color}
\usepackage{mathtools}
\usepackage{braket}
\usepackage{setspace}
\usepackage{graphicx}
\usepackage{hyperref}
\usepackage{starfont}
\usepackage{bbold}
\usepackage[shortlabels]{enumitem}
\usepackage{framed}
\usepackage{empheq}
\usepackage{amsmath}
\usepackage{tikz-feynman}
\usepackage{simpler-wick}
\usepackage{wasysym}
\usepackage{mhchem}
\usepackage{soul}

 \newcommand{\lsim}{{\;\raise0.3ex\hbox{$<$\kern-0.75em\raise-1.1ex\hbox{$\sim$}}\;}}
\newcommand{\gsim}{{\;\raise0.3ex\hbox{$>$\kern-0.75em\raise-1.1ex\hbox{$\sim$}}\;}}
\def\bea{\begin{eqnarray}}
\def\eea{\end{eqnarray}}
\def\bec{\begin{center}}
\def\ec{\end{center}}

\def\beq{\begin{equation}}
\def\eeq{\end{equation}}

\def\bea{\begin{eqnarray}}
\def\eea{\end{eqnarray}}
\def\beq#1\eeq{\begin{align}#1\end{align}}
\def\beqnn#1\eeq{\begin{align*}#1\end{align*}}
\def\ba{\begin{array}}
\def\ea{\end{array}}
\def\bc{\begin{center}}
\def\ec{\end{center}}
\def\nn{\nonumber}

\def\e{\epsilon}

\begin{document} 

\title{Cooling of young neutron stars and  dark gauge bosons}

\author{Deog Ki Hong} 
\email{dkhong@pusan.ac.kr}
\affiliation{Department of Physics,   Pusan National University, Busan 46241, South Korea}

\author{Chang Sub Shin} 
\email{csshin@ibs.re.kr}
\affiliation{Center for Theoretical Physics of the Universe, Institute for Basic Science (IBS), Daejeon, 34126, Korea}

\author{Seokhoon Yun} 
\email{SeokhoonYun@kias.re.kr}
\affiliation{Korea Institute for Advanced Study, Seoul 02455,  South Korea}
\affiliation{Dipartimento di Fisica e Astronomia, Universit\`a degli Studi di Padova, Via Marzolo 8, 35131 Padova, Italy}
\affiliation{Istituto Nazionale di Fisica Nucleare (INFN), Sezione di Padova, Via Marzolo 8, 35131 Padova, Italy}

\preprint{PNUTP-20/A03,\, CTPU-20-27,\, KIAS-P20069}

\begin{abstract}
 The standard cooling scenario in the presence of nucleon superfluidity fits rather well to the observation of the neutron stars.  It implies that the stellar cooling arguments could place a stringent constraint on the properties of novel particles.
We study in particular the cooling rate induced by dark gauge bosons for very young neutron stars: remnants of Cassiopeia A and SN1987A.  The cooling is dominantly contributed either by the nucleon pair breaking and formation in the core or by the electron bremsstrahlung in the crust, depending on the age of the stars and the form of the couplings. We compute  how much the cooling curve of the young neutron stars could be modified by the extra dark gauge boson emission and obtain the bound for the dark gauge boson when its mass is lower than $\mathcal{O}(0.1)\,{\rm MeV}$; for the dark photon we find the mixing parameter times its mass  $\varepsilon m_{\gamma^\prime} < 1.5 \times 10^{-8}\,{\rm MeV}$ and for the ${\rm U}(1)_{B-L}$ gauge boson its coupling to nucleons and electrons  $e^\prime < 5\times 10^{-13}$. 
We also discuss the possibility that the rapid cooling of Cas A might provide a hint for the existence of the ${\rm U}(1)_{B-L}$ gauge boson of mass around ${\rm eV}$ and its coupling $e^\prime \sim 10^{-13}$.
\end{abstract}

\maketitle

\section{Introduction}
\label{sec:Intro}

Neutron stars (NS) are one of the typical remnants in the core of supernova explosion and considered as the densest objects, directly observed in Nature.
Indeed, their core is constituted by the strongly-compressed nuclear (or even possibly quark) matter and the average density is estimated to be a few times the normal nuclear density, $\rho_0\equiv 2.8\times 10^{14}~{\rm g/cm^{3}}$, so that the average distance between nucleons is close to the size of nucleons $\sim1\,{\rm fm}$.

To understand the properties of NS, one needs to know  how elementary particles interact with each other in the extreme conditions at the most fundamental level.
In this sense, NS are good astrophysical laboratories searching for new physics beyond the Standard Model (SM), complementary to the laboratory searches as well as low density stellar objects~\cite{Raffelt:1996wa} like red giants and horizontal branch stars.
Young NS, whose ages are less than $10^4 \text{--}  10^5$ yr, could be attractive sources to detect a signal of new light particles like axions and light dark matter.  This is because the young NS still contain enough thermodynamic energy to become a factory for new particles, and there is little contamination from any other uncertain heating process.
Among many observations, there are particularly well isolated two young NS: the Cassiopeia A (Cas A) which is a supernova remnant in the Cassiopeia constellation, and a compact object in the remnant of SN1987A. Both are quite relevant to track the early stage of the thermal evolution of NS.

After Cas A was discovered by the first-light CHANDRA observation~\cite{Hughes:1999ph},  its surface temperature has been well measured to show a rapid steady decrease of about 2-4\% over 18 years from 2000~\cite{Heinke:2010cr,Wijngaarden:2019tht}\footnote{
Ref.~\cite{Posselt:2018xaf} pointed out that the data used in Ref.~\cite{Heinke:2010cr}  to estimate the cooling rate of the surface temperature may have been contaminated by certain instrumental effects. However, the upper limits on the temperature decrease provided in Ref.~\cite{Posselt:2018xaf}  are still consistent with the cooling rate analyzed by Refs.~\cite{Wijngaarden:2019tht,Ho:2019vbn}  based on more data.} using the non-magnetic partially ionized carbon atmosphere models of Ref.~\cite{Ho:2009mm} in order to fit the X-ray spectrum of Cas A.
The observed cooling rate, however, turns out to be too large to be accounted for the standard modified Urca process~\cite{Heinke:2010cr}, which is known to be the main cooling mechanism for young NS.
It is later argued that the rapid cooling may have been triggered by the enhanced emission of neutrinos from the NS core through the so-called Cooper-pair breaking and formation (PBF)~\cite{Page:2010aw,Shternin:2010qi,Ho:2014pta,Wijngaarden:2019tht}.
If further observations of Cas A confirm the cooling curve including the PBF process, that would be the first direct evidence of the superfluidity in NS, which has been predicted long ago~\cite{Migdal}. 

The another NS showing its early history is the conjectured compact object in the remnant of SN1987A, which is dubbed NS1987A. Through the recent observations attained by the Atacama Large Millimeter Array with its high angular resolution, it was identified from the infrared excess of a local dust blob near the predicted location of a kicked compact remnant~\cite{Cigan:2019shp}. This strongly suggests the presence of the neutron star as a cooling remnant~\cite{Page:2020gsx}. Interestingly, the expected luminosity fits well the standard cooling scenario~\cite{Page:2004fy,Yakovlev:2004iq,Page:2009fu} whose curve is consistent with the observations of the rapid cooling of Cas A  with the nucleon superfluidity~\cite{Page:2020gsx}.

In this paper, we provide new constraints on the existence of light dark particles with certain couplings,  based on the fact that the thermal histories of Cas A and NS1987A  fit well with the standard cooling curves. Recently, the rapid cooling of Cas A  was used to constrain the couplings of the QCD axion to find a bound comparable with that of SN1987A~\cite{Leinson:2014ioa,Hamaguchi:2018oqw}.
We consider  here other well-motivated hypothetical particles, namely  the dark photon~\cite{ArkaniHamed:2008qn} and the ${\rm U}(1)_{B-L}$ gauge boson, which we call collectively dark gauge bosons in this paper.  
While the dark photon couples to the electromagnetically charged particles with plasma suppression by the kinetic mixing~\cite{Holdom:1985ag}, the interaction of  the ${\rm U}(1)_{B-L}$ gauge boson to neutrons is not suppressed by the plasma effects.
Thus, the neutron superfluidity at around the age of Cas A is much more relevant for the production of the ${\rm U}(1)_{B-L}$ gauge bosons.
We note that the dark photon emission from the crust through the electron scattering could play the crucial role for the cooling of NS1987A, which leads to a bit more stringent bound on the dark photon compared to that from the rapid cooling of Cas A.
We discuss in detail such age dependence of dark gauge bosons in NS.

There are several studies on the constraints for dark photons from the sun, the HB stars and red giants~\cite{An:2013yfc,Redondo:2013lna,Hardy:2016kme}
and also from SN1987A~\cite{Chang:2016ntp}.
There are also interesting constraints on  ${\rm U}(1)_{B-L}$ gauge bosons from various studies such as the fifth force searches~\cite{Murata:2014nra}, the BBN~\cite{Heeck:2014zfa, Knapen:2017xzo}, the stellar cooling~\cite{Hardy:2016kme,An:2014twa,Knapen:2017xzo}. 
How our results on the dark photons and the ${\rm U}(1)_{B-L}$ gauge bosons fare with the previous studies is summarized in Fig.'s~6 and 7 in Section III, respectively.
We find that our results on the dark gauge bosons from the young NS (Cas A and NS1987A) are consistent with previous constraints, though there could be an interesting implication from Cas A for 
the ${\rm U}(1)_{B-L}$  gauge boson of mass around $1~{\rm eV}$ and its coupling, a few times $10^{-13}$.

The paper is organized as follows.
In Sec.~\ref{sec:DarkProduction}, we describe the physics of the dark gauge bosons in the dense (and exotic such as nucleon superfluidity) circumstance and derive its volume emissivity. In Sec.~\ref{sec:AnD} we analyze our findings for NS and discuss their consequences. We then provide our conclusion in Sec.~\ref{sec:con}.

\section{Dark gauge boson production in the neutron star}
\label{sec:DarkProduction}

Additional ${\rm U}(1)$ gauge symmetries beyond the SM are ubiquitous in string compactification as the unified theory of all forces.
A dark gauge boson ($A'_\mu$) can be a good candidate for dark matter or a messenger of the dark sector to the SM sector, depending on its mass and the couplings.  
Focusing on the interactions between $A'_\mu$ and the SM particles, we introduce a dark gauge boson as  the only new particle beyond the SM  to discuss its effects on  the NS cooling. 
The relevant effective Lagrangian below the electroweak scale is then given as
\bea
\mathcal{L}_{\rm eff} & = & -\frac{1}{4}F_{\mu\nu}F^{\mu\nu}  - \frac{1}{4}F^{\prime}_{\mu\nu}F^{\prime\mu\nu} 
-\frac{1}{2} m_{\gamma'}^2 A_\mu' A'^{\mu}  \nonumber \\
&& 
 +\frac{\varepsilon}{2}F_{\mu\nu}F^{\prime\mu\nu} + eA_\mu J_{\rm EM}^\mu  + e^\prime A^{\prime}_\mu J^{\prime \mu}   \, ,
 \label{eff_lag}
\eea
where $F_{\mu\nu} = \partial_\mu A_\nu - \partial_\nu A_\mu$, $F^\prime_{\mu\nu} = \partial_\mu A_\nu^\prime - \partial_\nu A_\mu^\prime$ denote the field strength  of the photon $A_\mu$, the dark gauge boson $A^\prime_\mu$ respectively. 
$m_\gamma'$ is the mass of $A'_\mu$ and 
$e J^\mu_{\rm EM}$ is the electromagnetic (EM) current. 
The (marginal) interaction between the dark gauge boson and the SM particles is shown in the second line of Eq.\,(\ref{eff_lag}); $\varepsilon$ is the dimensionless kinetic mixing parameter and 
$e' J^{\prime \mu}$ is the dark ${\rm U}(1)$ current comprising the SM fields.
In this paper, we consider the cases that the dark gauge boson couples only to the vector SM current, $J_\mu^\prime = \sum_{i=e,{\rm p},{\rm n}} q^\prime_i \bar{\psi}_i \gamma_\mu \psi_i$, where $q_i^{\prime}$ is the dark gauge charge of the SM fermion, $\psi_i$.

The properties of the additional gauge boson can be constrained from the astrophysical data if it is feebly interacting but copiously produced during the stellar evolution.  Inside the dense matter such as  NS,   the photons and the electromagnetic excitations like plasmons are screened and Landau-damped or Higgsed due to the medium effects of the charged particles. 
Because the medium effects are the collective phenomena of many particles, in general,  it is difficult to estimate them.  However, when the density is high enough as in NS, the hard-dense loop (HDL) approximation\,\cite{Blaizot:2000fc} is reliable. 
In the limit of the weak coupling, the response becomes linear and can be described by the polarization tensor 
\bea
\Pi^{\mu\nu}_{\gamma\gamma} = e^2 \left<J_{\rm EM}^\mu J_{\rm EM}^\nu\right> = \pi_{\rm T} \sum\epsilon_{\rm T}^\mu\epsilon_{\rm T}^\nu + \pi_{\rm L}\epsilon_{\rm L}^\mu\epsilon_{\rm L}^\nu \, .
\label{eq:Polarization}
\eea
For the external four-momentum $K = (\omega,\vec{k})$ in the rest frame of the medium, the longitudinal polarization vector is given by $\epsilon_{\rm L}^\mu = (k,\omega \vec{k}/k)/ \sqrt{ \omega^2-k^2}$, where $k=|\vec k|$. 
For $\vec k = k\hat z$, the transverse polarization vectors are taken as  $\epsilon_{\rm T}^\mu = (0,1,\pm i,0)/\sqrt{2}$. These can be easily generalized to the case with arbitrary directions of $\vec k$.\footnote{The formula for the polarization tensor Eq.~(\ref{eq:Polarization}) holds for any fluid, i.e. normal or superfluid. The fluid phase just changes the dispersion relation of photon not its polarization.}. 

In the Coulomb gauge, the effective propagators of the electromagnetic field are written as~\cite{Braaten:1993jw}
\bea
\left<A^i A^j\right> & = & \frac{1}{\omega^2 - k^2 - \Pi_{\rm T}}\left(\delta^{ij} - \frac{k^i k^j}{k^2}\right),
\label{trans} \\
\left<A^0 A^0\right> & = & \frac{1}{k^2 - \Pi_{\rm L}} \, .
\eea
As shown in Ref.~\cite{Braaten:1993jw}, the polarization functions  $\Pi_{\rm T,L}$ become in the leading order in the electromagnetic coupling constant $\alpha$ \footnote{In the proton superfluid the proton Cooper pair breaks the ${\rm U}(1)$ electromagnetism and gives rise to the Meissner mass to photons, the size of which is similar to the $\pi_{\rm T}$ in Eq.~\eqref{eq:TransversePlasmon}. Namely,  $\omega$ only in the transverse part, Eq.~(\ref{trans}),  is shifted by the proton gap~\cite{Hong:1999fh}. The Meissner mass of the photon is hence negligible in the dark gauge boson emissivity of the superfluid core, where the longitudinal polarization is dominant as shown later in this paper.}
\bea
\Pi_{\rm T} & = & \omega_{\rm P}^2\left[1+\frac{1}{2}G\left(v_{*}^2k^2/\omega^2\right)\right] \equiv \pi_{\rm T}  
\label{eq:TransversePlasmon}, \\ 
\Pi_{\rm L} & = & \omega_{\rm P}^2\frac{k^2}{\omega^2}\frac{1-G\left(v_{*}^2k^2/\omega^2\right)}{1-v_{*}^2k^2/\omega^2} \equiv \frac{k^2}{\omega^2 -k^2}\pi_{\rm L} \, ,
\label{eq:LongitudinalPlasmon}
\eea
where $\omega_{\rm P}$ is the plasma frequency, $v_*$ denotes the typical electron velocity in the medium, and  $G(x)$ is given as~\cite{Braaten:1993jw} 
\bea
G(x) =\frac{3}{x}\left(1- \frac{2x}{3} - \frac{1-x}{2\sqrt{x}} \ln\frac{1+\sqrt{x}}{1-\sqrt{x}}\right).
\eea 
Since NS are dense and cold enough,  the Fermi momentum of the electron in NS, $|\vec p_{F, e}|$ ($=\mathcal{O}(1)\,{\rm fm}^{-1} =\mathcal{O}(100)\,{\rm MeV}$),  is much greater than its mass ($m_e$) and its temperature $T$ ($\leq \mathcal{O}(10^9)\,{\rm K} = \mathcal{O}(0.1)\,{\rm MeV}$). The electrons are therefore highly relativistic, $1 - v_* = \mathcal{O}(m_e^2/|\vec p_{F,e}|^2)$. 
In this circumstance, the plasma frequency is estimated to be
\bea 
\omega_{\rm P} = \left(\frac{4 \pi \alpha n_e }{E_{F,e}}\right)^{1/2} =\left(\frac{|\vec p_{F,e}|}{1/{\rm fm}}\right) \mathcal{O} (10)\,{\rm MeV} .\eea
Taking the external momentum to be on mass-shell of the dark gauge bosons, i.e. $\omega^2-k^2= m_{\gamma'}^2$, we can approximate
\bea 
\pi_{\rm T} \simeq \frac{3}{2} \omega_{\rm P}^2,\quad
\pi_{\rm L} \simeq 3\omega_{\rm P}^2 \frac{m_{\gamma^\prime}^2}{T^2} \ln \left(\min\left[\frac{|\vec p_{F,e}|}{m_e}, \frac{T}{m_{\gamma^\prime}}\right]\right) \label{eq:Approximation}\eea 
in the limit of $T\gg m_{\gamma'}$.

On one hand, when the dark current  $e'J'^\mu$ contains the electrically charged particles, especially the electrons, 
there is the in-medium mixing between $A'_\mu$ and $A_\mu$~\cite{An:2013yfc,Redondo:2013lna,Redondo:2008ec,Hardy:2016kme}.
This effect is simply given by~\cite{Hardy:2016kme}
\bea
\Pi^{\mu\nu}_{\gamma\gamma^\prime} = e e^\prime \left<J_{\rm EM}^\mu J'^\nu\right> \approx (e^\prime q_e^\prime/eq_e) \Pi^{\mu\nu}_{\gamma\gamma} \, ,
\eea
where $q_e^\prime$ and $q_e ( = -1)$ are the dark and EM gauge quantum numbers of the electrons in the unit of $e'$ and $e$, respectively, and the second expression is derived in the lowest order in couplings, $e$ and $e^{\prime}$.

\begin{table}[t!]
\centering
\begin{tabular}{|c|c|c|c|}
\hline \rule{0pt}{13pt}  
Model &   $e_{\rm eff}^{\rm e}$ & $ e_{\rm eff}^{\rm p} $ & $\ e_{\rm eff}^{\rm n}\ $ \\[1ex] 
 \hline\hline 
 \rule{0pt}{13pt}
 $\textrm{Dark photon}$ \qquad  & $\varepsilon e  m_{\gamma^\prime}^2/\pi_{\rm T,L}$ & $-\varepsilon e m_{\gamma^\prime}^2/\pi_{\rm T,L} $ & $\text{-}$ \\[1ex]  \hline 
\rule{0pt}{13pt}  ${\rm U}(1)_{B-L}$ &  $e^\prime m_{\gamma^\prime}^2/\pi_{\rm T,L}$ &$-e^\prime m_{\gamma^\prime}^2/{\pi_{\rm T,L}}$ & $e^\prime$ \\[1ex] 
\hline
\end{tabular}
\caption{The effective couplings ($e^f_{\rm eff}$ defined in Eq.~(\ref{eq:MatrixElement})) to the currents of the electron, the proton, and the neutron for the given dark gauge boson scenario: dark photon,  ${\rm U}(1)_{B-L}$. The values are estimated in the limit of the dense medium (i.e. $\pi_{\rm T,L} \gg T^2 > m_{\gamma^\prime}^2$)}
\label{tab:EffectiveCouplings}
\end{table}

Taking into account the above considerations,  we evaluate the matrix element for the production of dark gauge bosons inside 
the dense matter as
\bea 
\hskip -0.4cm \mathcal{M}_{\rm T,L} & = & \mathcal{M}_{\rm T,L}^{J^\prime}+\mathcal{M}_{\rm T,L}^\varepsilon + \mathcal{M}_{\rm T,L}^{\Pi_{\gamma\gamma^\prime}} \equiv e_{\rm eff}^f  [j_\mu^f]\epsilon_{{\rm T,L}}^\mu \, ,
\label{eq:MatrixElement}
\eea 
where
\bea
\mathcal{M}^{J^\prime}_{\rm T, L} & = & e^\prime q_f^\prime   [j_\mu^f]\epsilon^\mu_{\rm T,L} \, , \\
\mathcal{M}^{\varepsilon}_{\rm T, L} & = & \left(\frac{\varepsilon m_{\gamma^\prime}^2}{m_{\gamma^\prime}^2-\pi_{\rm T,L}}\right)  e q_f [j_\mu^f]\epsilon^\mu_{\rm T,L} \, , \\
\mathcal{M}^{\Pi_{\gamma\gamma^\prime}}_{\rm T, L} & = & \left( \frac{e' q_e^\prime }{e q_e}\frac{ \pi_{\rm T,L}}{m_{\gamma^\prime}^2-\pi_{\rm T,L}}\right)   e q_f [j_\mu^f]\epsilon^\mu_{\rm T,L} \, , 
\eea 
and 
$[j_\mu^f] \equiv \int d^4 x\, e^{ i K\cdot x} \langle f_{\rm fin} | \bar\psi_f \gamma_\mu\psi_f | f_{\rm in} \rangle $ 
for the initial (in) and final (fin) fermions, $f=e, {\rm p}, {\rm n}$, of the current.
Each superscript in Eq.~\ref{eq:MatrixElement} stands for the contribution from the dark ${\rm U}(1)$ current ($J^{\prime}$), the kinetic mixing 
($\varepsilon$), and the in-medium mixing ($\Pi_{\gamma\gamma^\prime}$).
Here, $e_{\rm eff}^f$ denotes the effective coupling between $A^\prime_\mu$ and the SM fermions ($f$) in the medium, 
\bea
e_{\rm eff}^f =  e^\prime \left( q_e^\prime q_f - q_f^\prime q_e\right) + 
  \left(\varepsilon e - e^\prime q_e^\prime\right)q_f \frac{ m_{\gamma^\prime}^2}{m_{\gamma^\prime}^2 - \pi_{\rm T,L}} \, .
\label{eq:EffectiveCoupling}
\eea
Notice that the in-medium effect comes mainly from the scattering of the electrons, being the lightest charged particle.   If $q_f^\prime/q_e'$ is equal to  $q_f/q_e$, the first term in the bracket of Eq.~\eqref{eq:EffectiveCoupling} vanishes so that the effective coupling for a given $f$ in such a dense medium is suppressed by a factor of $m_{\gamma^\prime}^2/\pi_{\rm T,L}$~\cite{Hardy:2016kme}.

In this paper, we consider two benchmark models for the dark gauge boson: the dark photon which couples to the SM particles only through the dimensionless kinetic mixing $\varepsilon$, and the ${\rm U}(1)_{B-L}$ gauge boson without the kinetic mixing with the photon.
The emission of dark gauge bosons from the currents of the electrons and the protons in medium is suppressed for both models by the plasma effect~\cite{Hardy:2016kme}, while the dark gauge boson productions from the neutron currents are drastically different between two models. 
In the dark photon scenario, there is almost no emission from the neutron currents, because neutrons are electrically neutral. However, in the ${\rm U}(1)_{B-L}$ case, dark gauge bosons are dominantly emitted through the neutrons because the plasma suppression is negligible for low temperatures where the neutrons are effectively structureless. The effective couplings in the each scenario are summarized in  Table~\ref{tab:EffectiveCouplings}.

Now, let us calculate the dark gauge boson production rate inside NS for a given temperature. 
The matter is in an exotic circumstance: a relatively low temperature with a very high nucleon density leading to a condensed phase as superfluidity (for neutrons) or superconductivity (for protons). 
In such environments, 
we find that the following two dark-matter emission processes  are important for the NS cooling: (i) the nucleon pair breaking and formation (PBF) in the core and (ii) the electron bremsstrahlung, interacting with heavy nuclei in the crust.
The other processes such as nucleon bremsstrahlung in the core are subdominant.

\subsection{Emission through the  breaking and  formation of Cooper pairs }
\begin{figure}[ht!]
\centering
\begin{minipage}{0.48\textwidth}
\centering
  \includegraphics[width=.48\linewidth]{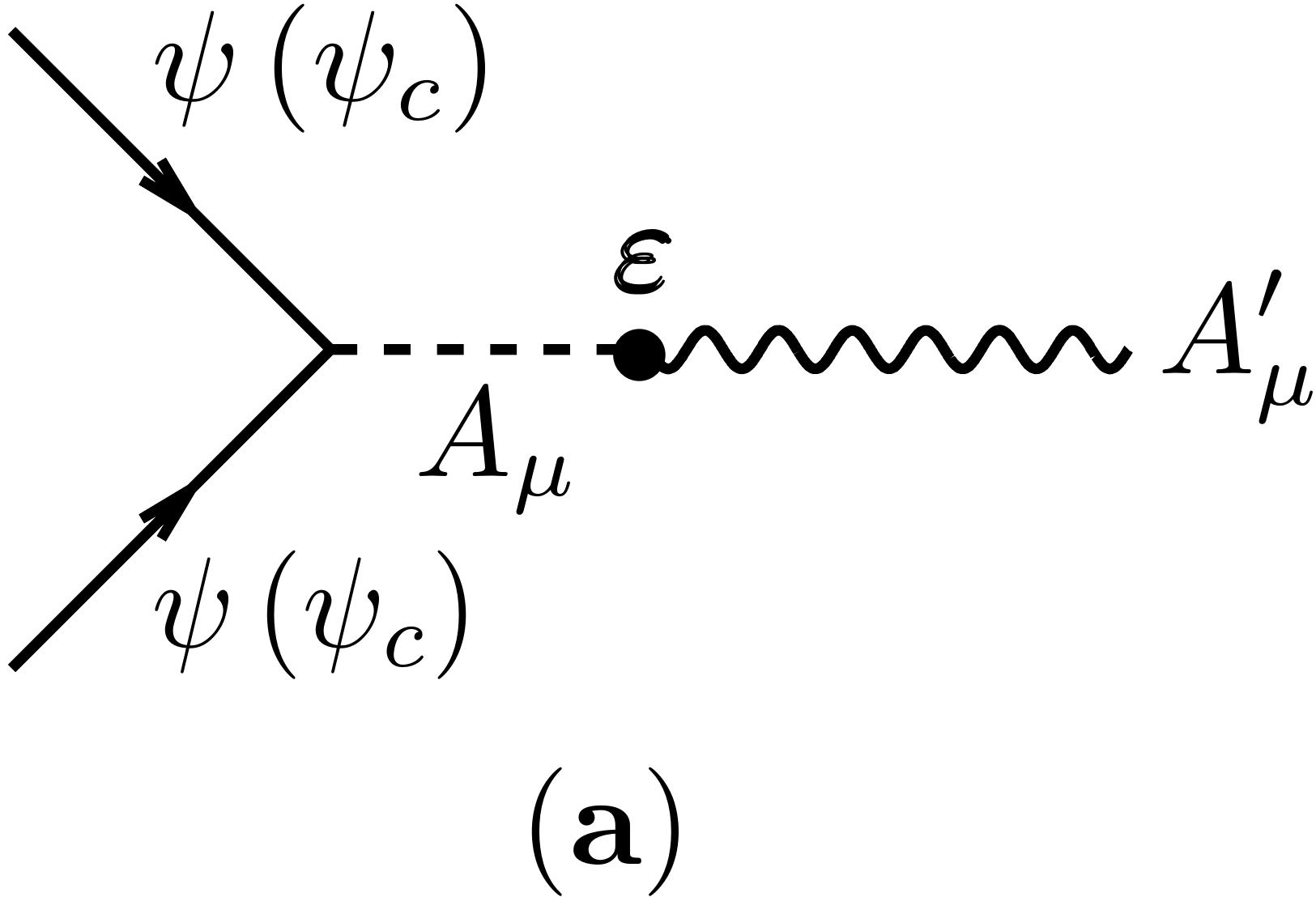}
\qquad
  \includegraphics[width=.35\linewidth]{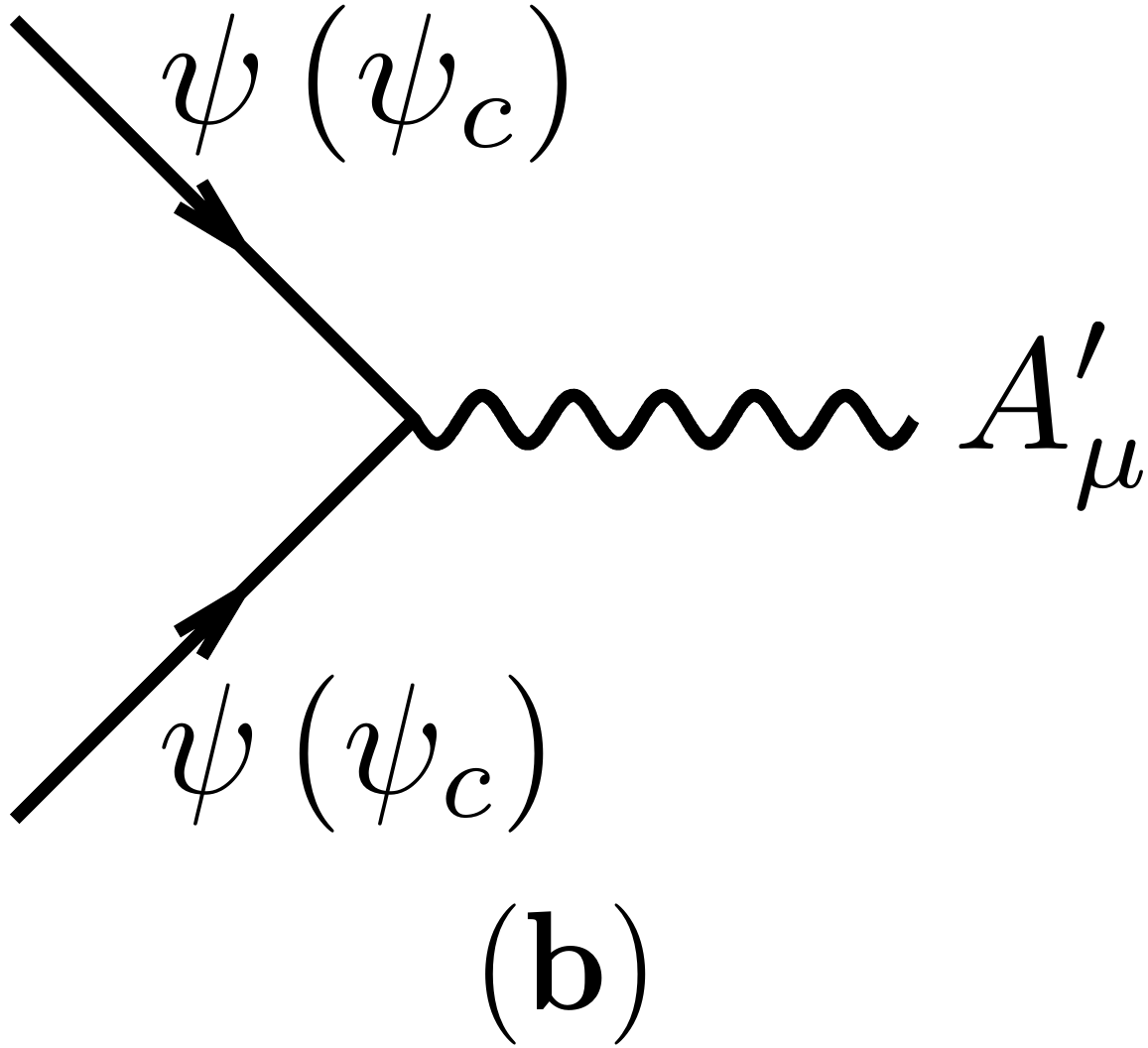}
\end{minipage}%
\caption{(a) Dark photons mix with  photons $A_{\mu}$  in the proton Cooper-pair  (or hole-pair) formation. The bullet denotes the mixing with the kinetic mixing parameter $\varepsilon$. 
(b) ${\rm U}(1)_{B-L}$ gauge boson emission by the neutron triplet Cooper-pair (or hole-pair) formation. }
\label{fig1}
\end{figure}

Once NS cool enough,  the superfluid and superconducting phase transitions occur at temperatures of $0.1-1\,{\rm MeV}$ (a fraction of $10^{10}\,{\rm K}$) 
from the Bardeen-Cooper-Schriefer (BCS) pairing among nucleons~\cite{Flowers:1976ux}.   
Since the nuclear interactions are known to be repulsive at short distances but attractive at long distances, 
the structure of the phase transition depends on the nucleon density. 
Including the medium effects, detailed analysis shows that  p-wave pairing is preferred at high  core densities ($\rho>\rho_0$),  but at low densities ($\rho\lesssim\rho_0$), s-wave pairing is more preferred. 
 In the core of NS, the neutron density is quite high as $5 - 10\, \rho_0$, whereas the protons are less densely packed, so  it is expected that the neutrons form p-wave pairs to become superfluid and the protons become superconductor by forming s-wave pairs. 
In NS the gap of neutron p-wave paring  ($\Delta_{{\rm n}^3\! P_2}\sim 0.1~{\rm MeV}$) is generically smaller than or comparable to that of the proton s-wave pairing ($\Delta_{{\rm p}^1\!  S_0}\sim 0.1-0.5~{\rm MeV}$). Therefore, as the temperature decreases, the superconducting phase transition occurs earlier than the neutron superfluid phase transition~\cite{Lombardo:2000ec,Dean:2002zx}. 
The critical temperatures ($T_c$) is in general proportional to its gap $T_c =c \Delta(0)$, where $c=0.57$ for the BCS model but the explicit value of $c$ varies with ${\cal O}(0.1-1)$, depending on models.

The modes near the Fermi surface are relevant for the NS cooling, and  their dynamics is conveniently described by the high density effective theory (HDET)~\cite{Hong:1998tn,Hong:1999ru} as following
\bea
{\cal L}_{\rm eff}& = &\sum_{\vec v_F}\left[\bar\psi(\vec v_F,x)\gamma^0V^{\mu}\partial_{\mu}\psi(\vec v_F,x) \right.  \nn \\
&& \! \! \quad  \left. -\left(\Delta \bar{\psi}(\vec v_F,x)\gamma^5 \psi_c(\vec v_F,x) + {\rm h.c.}\right) \right]+\cdots \, , \quad
\eea
where $V^{\mu}=(1,\vec v_F)$ with $\vec v_F$ being the Fermi velocity, $\Delta$ is the paring gap and $\psi$ and $\psi_c \equiv C\bar{\psi}^{\rm T} = i\gamma^2 \psi^*$ denote the particle and the hole state, respectively, that carry the residual energy and momentum around the Fermi surface. The ellipsis denotes the gauge interactions of the modes with photons, dark ${\rm U}(1)$ gauge bosons (e.g. dark photon or ${\rm U}(1)_{B-L}$ gauge boson in this paper) and higher order interactions among themselves.   At temperatures close to but below the critical temperatures,  the dominant cooling process is through the pair breaking and formation of nucleons into particles $X$ such as dark photons or ${\rm U}(1)_{B-L}$ gauge bosons: 
\begin{equation}
\psi+\psi \to X\, \quad \text{or} \, \quad  \psi\to \psi_c+X\, .
\end{equation}
Since the pair-formation and pair-breaking processes are equilibrated, we just consider the pair-formation process,  emitting dark photons (Fig.~\ref{fig1}\,a)  and  ${\rm U}(1)_{B-L}$ gauge bosons (Fig.~\ref{fig1}\,b)\footnote{There are also the emitted ${\rm U}(1)_{B-L}$ gauge bosons through the proton-singlet and neutron-singlet PBF but their rates are negligible compared to the neutron triplet PBF process}.

The dark gauge boson emissivity can be written as 
\begin{equation}
Q_V^{\rm PBF}=2 \int\frac{{{\rm d}^3\vec k}}{{2\omega (2\pi)^3}}{\rm d}W_{i\to f}\omega f_F\left(\frac{\epsilon_p}{T}\right)f_F\left(\frac{\epsilon_{p'}}{T}\right)\,,
\end{equation}
where $\omega$ is the energy of the dark gauge boson carrying the momentum $\vec k$, 
$\epsilon_p=(\Delta^2+|\vec v_F|^2 \left(|\vec p|- |\vec p_F|\right)^2)^{1/2}$ is the the energy of the quasi particle with the momentum $\vec p$ and the Fermi-velocity $\vec v_F$, 
$\epsilon_{p'}$ is the energy of the quasi particle with the momentum $\vec p'$ and the opposite Fermi-velocity $-\vec v_F$,  $f_F(x)=(e^x + 1)^{-1}$ is the Fermi-Dirac distribution of the quasi particles. 
The transition rate is given as
\bea
{\rm d}W_{i\to f} & = & \frac{{\rm d}^3\vec p}{(2\pi)^3}\frac{{\rm d}^3\vec p^{\prime}}{(2\pi)^3}\left|M\right|^2 \nn \\
&& \times(2\pi)^3\delta(\epsilon_p+\epsilon_{p^{\prime}}-\omega)\delta^3(\vec p+\vec p^{\prime}-\vec k)
\, ,
\eea
with $M = \mathcal{M}_\mu \epsilon^\mu$, the matrix element of the PBF process. $\epsilon^\mu$ is the polarization vector of the dark gauge boson, and $\mathcal{M}_\mu = e_{\rm eff}^{\rm N} \left<J_\mu^{\rm N}\right>$ denotes the matrix element of the corresponding vector current of each nucleon ${\rm N} = {\rm n, p}$.

As noted in~\cite{Leinson:2006gf}, in order to satisfy the conserved vector current hypothesis in weak interactions, the vertex for the vector currents has to be modified in superfluid to include the supercurrent mode.
Namely, the vector current has additional contributions from the supercurrent modes which mix with the external vector fields as $J_{\mu}=\bar\psi\gamma_{\mu}\psi+if_{\phi}\partial_{\mu}\phi$.
As a consequence, the matrix element is modified by the collective corrections as~\cite{Leinson:2006gf} \footnote{As noted in Ref.~\cite{Kolomeitsev:2008mc}, the approach of Ref.~\cite{Leinson:2006gf} is valid only in the limit $\omega \ll \Delta$. Analogous to the neutrino case, however, the nucleon PBF emission rates of  dark gauge bosons, based on Ref.~\cite{Leinson:2006gf}, slightly differ from the more general study~\cite{Kolomeitsev:2008mc}. Explicitly, we find that the proton singlet PBF rate is $4/5$ times smaller but the neutron triplet PBF rate is same, if one follows the analysis done in Ref.~\cite{Kolomeitsev:2008mc}.}
\bea
&&i\mathcal{M}_0 =\frac{\Delta}{\e_p} \left( \frac{  v_F^2k^2}{3\,\omega^2}\right),\quad
i\mathcal{M}_\| =  \frac{\Delta}{\e_p} \left(  \frac{ v_F^2 k}{3\,\omega}\right),\nonumber\\
&&i\mathcal{M}_\bot = - \frac{\Delta}{\e_p} \left( \frac{  v_F^2 k}{2\,\e_p} \cos\theta \sin\theta\right) \, ,
\eea
where $\mathcal{M}_0$ is the temporal component of the matrix element and $\mathcal{M}_\|$ and $\mathcal{M}_\bot$ is the spatial component of the longitudinal and the transverse part of the matrix element, respectively, and $\cos\theta = \hat{k}\cdot \hat{p}$.

In both of the dark photon and the ${\rm U}(1)_{B-L}$ scenario, the dark gauge bosons could be produced by the proton singlet (${\rm p}^1\!S_0$) PBF.
Considering that the effective coupling to the proton is suppressed by the plasma effect, one can easily realize that the longitudinal polarization of such gauge bosons is dominantly produced in a typical case.
Integrating over the phase space with the HDL approximation, we get the volume emissivity of the ${\rm p}^1\!S_0$ PBF process
\bea
Q_{\gamma^\prime, {\rm L}}^{{\rm p}^1\! S_0}  & \simeq &   g'^2 m_{\rm p}^*   |\vec p_{F,{\rm p}}|  T^3 \!\left(\frac{ m_{\gamma^\prime}^6  }{\pi_{\rm L}^2 T^2} \right) \!\!
\left(\frac{8 v_{F, {\rm p}}^4 F_3 \left(z_{{\rm p}^1\! S_0}\right) }{9\,\pi^3} \right)\,\, \, \,\,\,\,  \nn\\
& \simeq & 2.56 \times 10^{17} \, {\rm erg}\,{\rm cm}^{-3}{\rm s}^{-1} \, \left(\frac{v_{F, {\rm p}}}{0.1}\right)^5 \left(\frac{m_p^*}{m_p}\right)^2   \nn \\
&& \times \left(\frac{1/{\rm fm}}{|\vec p_{F,e}|}\right)^{4} \left(\frac{10}{3 \ln\left(\min\left[\frac{|\vec p_{F,e}|}{m_e}, \frac{T}{m_{\gamma^\prime}}\right]\right)}\right)^2 \nn \\
&& \times \left(\frac{g^\prime m_{\gamma^\prime}}{10^{-9}\,{\rm MeV}}\right)^2 \left(\frac{T}{10^{9}\,{\rm K}}\right)^5 F_3 \left(z_{{\rm p}^1\! S_0}\right) 
\eea 
where $\pi_L$ is given by Eq.~(\ref{eq:Approximation}), $m_{\rm p}^*$ is the effective proton mass, $g'= \varepsilon e$ for the dark photon scenario, and $g'=e'$ for the ${\rm U}(1)_{B-L}$ gauge boson scenario.  
$F_3(z_{{\rm p}^1\! S_0})$ is defined by
\bea
F_n \left(z\right) = \int \frac{{\rm d}\Omega}{4\pi} \int_1^\infty {\rm d}y\, \frac{  z^{n+2} y^n}{\sqrt{y^2-1}}f_F\left(zy\right)^2 
\label{eq:F_n} 
\eea
with $n=3$ and $z= \Delta_{{\rm p}^1\! S_0}/T$, and can be approximated by~\cite{Sedrakian:2015krq}
\bea
F_3(z) \simeq \left(a_3 z^2 + b_3 z^4\right) \sqrt{1+ f_3 z} e^{-\sqrt{4z^2+h_3^2}+h_3} \, ,
\eea
where $a_3 = 0.158151, b_3 = 0.543166, h_3 = 0.0535359$, and $f_3 = \pi/4b_3^2$.

While  dark photons are  effectively not emitted by neutrons, ${\rm U}(1)_{B-L}$ gauge bosons do get emitted by the neutron currents directly.  Because of the larger number of the neutrons in the core, ${\rm U}(1)_{B-L}$ gauge bosons could be produced more actively, and their volume emission becomes an important source of the NS cooling.
Following the same procedure done previously for the ${\rm p}^1\!S_0$ PBF process, the volume emissivity through the ${\rm n}^3\!P_2$ PBF is given by
\bea
Q_{B-L}^{{\rm n}^3\! P_2}  & \simeq &  e'^2 m_{\rm n}^* |\vec p_{F, {\rm n}}| T^3 
\left(\frac{4 v_{F,{\rm n}}^4   F_1 \left(z_{{\rm n}^3\! P_2}\right)}{15\pi^3}\right)  \, , \nn \\
& \simeq & 1.53 \times 10^{17} \, {\rm erg}\,{\rm cm}^{-3}{\rm s}^{-1} \, \left(\frac{v_{F, {\rm n}}}{0.1}\right)^5 \left(\frac{m_{\rm n}^*}{m_{\rm n}}\right)^2 \nn\\
&& \times \left(\frac{e^\prime}{10^{-13}}\right)^2 \left(\frac{T}{10^{9}\,{\rm K}}\right)^3 F_1 \left(z_{{\rm n}^3\! P_2}\right) 
\eea
where $m_{\rm n}^*$ is the effective neutron mass, and $z_{{\rm n}^3\! P_2} = \Delta_{{\rm n}^3\! P_2}/T$.
Here the triplet gap depends on the angle between the quasi-particle momentum and the arbitrary given quantization axis, i.e. $\Delta_{{\rm n}^3\! P_2}^2 (T,\theta_{\rm n}) = \Delta_{{\rm n}^3\! P_2}^2 (T) {\cal F}(\theta_{\rm n})$ where ${\cal F}(\theta_{\rm n}) =  1+3\cos^2\theta_{\rm n}$ for orbital quantum number $m_j=0$ and $\sin^2\theta_{\rm n}$ for $m_j=\pm 2$ with the corresponding angle $\theta_{\rm n}$. 
$F_1 \left(z_{{\rm n}^3\! P_2}\right)$ is defined as $F_{n=1}(z=\Delta(T,\theta_{\rm n})/T)$ for Eq.~(\ref{eq:F_n}), also can be approximated by
\bea
\hskip -0.5cm F_1 \left(z = \bar{z}\sqrt{{\cal F}}\right) & \simeq  & \frac{\left(a_1 \bar{z}^2 + b_1 \bar{z}^4 + c_1 \bar{z}^6\right) \sqrt{1+ f_1 \bar{z}}}{\left(1+ d_1 \bar{z}^2\right) e^{\sqrt{4\bar{z}^2+h_1^2}-h_1}}  \, ,
\eea
where $\bar{z} \equiv \Delta_{{\rm n}^3\! P_2} (T)/T$, $a_1 = 0.363127$, $b_1 = 0.0369455$, $c_1=0.0000606479$, $d_1=0.19057$, $h_1 = 0.880488$, and $f_1 = a_1/4\pi$.

\subsection{Bremsstrahlung emission in the crust}

The crust of NS where $\rho < 0.5 \rho_0 \simeq 1.4 \times 10^{14}\, {\rm g}/{\rm cm}^3$ is mainly composed by relativistic electrons and heavy nuclei characterized by their charge $Z$ and atomic weight $A$. There are also dripped neutrons when $\rho > 4 \times 10^{11}\, {\rm g}/{\rm cm}^3$.
The important feature of the crust is that such heavy ions can be in a solidified lattice state (i.e. a Coulomb crystal) when~\cite{Slattery:1980zz} 
\bea
\frac{Z^2 \alpha}{a T} = 0.23 Z^2 \left(\frac{10^8\,{\rm K}}{T}\right)\left(\frac{\rho/A}{10^6\, {\rm g}/{\rm cm}^{3}}\right)^{1/3} > 178 \, ,
\label{eq:Melt}
\eea
where $a = (3/4\pi n_i)^{1/3}$ denotes the ion-sphere radius. 
Otherwise the nuclei form a Coulomb liquid state.

The dark gauge boson production inside the volume of the crust is dominated by the electron bremsstrahlung.
Since the inner crust typically satisfies the condition of Eq.~\eqref{eq:Melt}, we have to take into account the electron scattering on the crystalline lattice of ions~\cite{Itoh:1984xy,Pethick:1993rr,Pethick:1996yj} (or phonons~\cite{Itoh:1984yz,Itoh:1989yz,Yakovlev:1996fs}). 
A free electron with the wavenumber $\vec p$ is strongly mixed with the another free electron with wavenumber $\vec p+\vec K$ where $\vec K$ corresponds to a reciprocal lattice vector.
The periodic potential of the crystallized ions then causes a band structure in the dispersion relation of electrons with energy splitting~\cite{Pethick:1993rr,Pethick:1996yj}.
Consequently, the electron bremsstrahlung process in collisions with atomic nuclei occurs through 
the electron transition from an upper band state to a lower one.

Now let us derive the volume emissivity of the dark gauge boson bremsstrahlung through electrons moving in the crystalline lattice of ions in the crust.
The phonon contribution is neglected.
The dark gauge boson emission rate can be calculated as
\bea
Q_{\gamma^\prime}^{\rm eZ} = \sum_{\vec K} \int \frac{{\rm d}^3 \vec k }{2\omega(2\pi)^3} \frac{{\rm d}^3 \vec p_1}{(2\pi)^3}\frac{{\rm d}^3 \vec p_2}{(2\pi)^3} f_{F1}\left(1-f_{F2}\right) \omega \left| \mathcal{M}\right|^2 \nn \\
\times \left(2\pi\right)^4\delta(E_{\vec p_1}^+ - E_{\vec p_2}^- - \omega)\delta^{(3)}\left(\vec p_1 - \vec p_2 - \vec k\right),~~
\eea
where $\vec p_1$ and $\vec p_2$ are the spatial momentum of the initial (upper band) and final (lower band) electron, respectively.
Here the energy eigenvalues are given by $E^\pm_{\vec p} = (E_{\vec p} + E_{\vec p-\vec K})/2 \pm \mathcal{E}_{\vec p}$,  where $E_{\vec p}$ is the energy of a free electron, $\mathcal{E}_{\vec p} = (\xi_{\vec p}^2 + V_{\vec K}^2)^{1/2}$ with $\xi_{\vec p} = (E_{\vec p} - E_{\vec p-\vec K})/2$. 
The matrix elements of the lattice potential is given by
\bea
V_{\vec K} =  - \frac{ 4\pi^2 \alpha n_e v_\bot F_{\vec K} e^{-W}   }{\pi |\vec K|^2+4\alpha|\vec p_{F,{\rm e}}|^2}
\eea
with $v_\bot = (1-v_\|^2)^{1/2}=(1- |\vec K|^2/(2 |\vec p_{F, {\rm e}}|)^2)^{1/2}$, $F_{\vec K}$ the form factor for the charge distribution and $W$ the Debye-Waller factor from the vibrations of the ionic lattice which we shall neglect as in Ref.~\cite{Pethick:1996yj}.
Evaluating the scattering amplitude includes the knowledge of the electron band structure effect and sum over reciprocal lattice vectors $\vec K$.
For more details, one can refer to Ref.~\cite{Pethick:1993rr,Pethick:1996yj,Yakovlev:1996fs}.

Since the dark gauge boson emission in the crust is primarily through the electron current, its rate becomes always suppressed by the plasma effect as discussed antecedently.
Noting that the emission of the longitudinal component is dominant, the matrix element is given by
\bea
\left|\mathcal{M}\right|^2 & \approx & \left(e_{\rm eff}^{\rm e}\right)^2_L \epsilon_L^\mu \epsilon_L^\nu \sum_{\sigma_1, \sigma_2} \left<J_\mu\right>\left<J_\nu\right> \nonumber \\
& \approx &  \left( e_{\rm eff}^{\rm e}\right)^2_L \frac{m_{\gamma^\prime}^2}{\omega^2} \left(u_{\vec p_1}v_{\vec p_2} -v_{\vec p_1}u_{\vec p_2}\right)^2 \, .
\eea
In the expression, 
$( e_{\rm eff}^{\rm e})_{\rm L}$ denotes the effective coupling to the electron of the longitudinal component,   $u_{\vec p}= V_{\vec K}/\sqrt{2 \mathcal{E}_{\vec p} (\mathcal{E}_{\vec p} - \xi_{\vec p})}$ and $v_{\vec p} =  \sqrt{(\mathcal{E}_{\vec p} - \xi_{\vec p})/2 \mathcal{E}_{\vec p}}$.
Following the steps shown in Ref.~\cite{Pethick:1996yj}, we calculate the dark gauge boson emissivity through the electron bremsstrahlung process in the crust given by
\bea
Q_{\gamma^\prime}^{\rm eZ} & \simeq &  g'^2  |\vec p_{F, {\rm e}}| T^4 \left( \frac{m_{\gamma'}^6}{16\pi^4  T^2 \pi_L^2 } \right)  \sum_{\vec K} G\left(v_\| , t\right)  \nn \\
 & \simeq & 2.44 \times 10^{22} \, {\rm erg}\,{\rm cm}^{-3}{\rm s}^{-1} \,  e^{-\frac{|\vec p_{F,e}|}{0.13/{\rm fm}}}  \nn  \\
&& \times  \left(\frac{0.4/{\rm fm}}{|\vec p_{F,e}|}\right)^{1.8} \left(\frac{10}{3 \ln\left(\min\left[\frac{|\vec p_{F,e}|}{m_e}, \frac{T}{m_{\gamma^\prime}}\right]\right)}\right)^2 \nn \\
&& \times \left(\frac{g^\prime m_{\gamma^\prime}}{10^{-9}\,{\rm MeV}}\right)^2 \left(\frac{T}{10^{9}\,{\rm K}}\right)^{5.2} e^{-\frac{8.5\times 10^6\,{\rm K}}{T}}
\eea
for $ t \equiv \left| V_{\vec K}\right|/T$ and 
\bea
G\left(v_\| , t\right) & = & \frac{1}{2} \int_0^\infty {\rm d}x_1 \int_0^\infty {\rm d}x_2 \int_{x_{-}}^{x_{+}} {\rm d}x_3  \nn \\
&& \quad\frac{\varsigma^3}{\exp[\varsigma]-1} \left(1-\frac{\kappa_1 \kappa_2}{e_1e_2} - \frac{t^2}{e_1e_2}\right)
\eea
with $\kappa_{1,2} = v_{\|}\left| x_1 \pm x_2/2\right|$, $e_{1,2} = (t^2 +\kappa_{1,2}^2)^{1/2}$, $x_{\pm} = (v_\bot(e_1 + e_2) \pm ((e_1+e_2)^2 - v_\|^2 x_2^2)^{1/2})/v_\|^2$, and $\varsigma = v_\bot x_3 + e_1 + e_2$.

There are two important limiting cases: (i) low- and  (ii) high-temperature limit compared to the lattice potential $V_{\vec K}$.
In the low-temperature limit ($T \ll V_{\vec K}$), the dark gauge boson emissivity would decrease exponentially because the band gap is too big to excite electrons from the lower band to the upper one.
Since a smaller reciprocal vector possesses a larger gap potential and the reciprocal vector becomes smaller as NS cools, such contribution to the emissivity gets suppressed as the temperature decreases.  
In the high temperature limit ($T\gsim V_{\vec K}$), we find that each lattice reciprocal vector contribution to the dark gauge boson production is proportional to $(V_{\vec K}/T)^2$.
In other words, the emissivity from a specific reciprocal lattice vector becomes more and more efficient as cooling goes on until $T\approx V_{\vec K}$.
After summing all contributions from different reciprocal vectors, it is found that the temperature dependence of 
the dark gauge boson emissivity in the crust is approximately given as $T^6$ in most situations we are considering.

Before closing this section, 
let us discuss the effect of the emissions through the nucleon bremsstrahlung in the core.
In order to estimate the dark gauge boson emissivity through the ${\rm nn}$- or ${\rm pp}$-bremsstrahlung in core, we follow the steps shown in the Ref.~\cite{Iwamoto:1992jp} and apply it to the dark gauge boson.
The result of the emissivities is given by
\bea
Q_{\gamma^\prime}^{\rm nn} & = & \frac{5}{3}\frac{31}{3780\pi} e^{\prime 2} T^6  \left(\frac{f}{m_\pi}\right)^4 \left(m_{\rm n}^*\right)^2 |\vec p_{F,{\rm n}}| F\left(x_{\rm n}\right)\,\,\,\,\, \nn \\ 
& \simeq & 2.94\times 10^{12} \, {\rm erg}\,{\rm cm}^{-3}{\rm s}^{-1} \left(\frac{m_{\rm n}^*}{m_{\rm n}}\right)^2 \left(\frac{|\vec p_{F,{\rm n}}|}{1/{\rm fm}}\right) \nn \\
&& \times \left(\frac{e^\prime}{10^{-13}}\right)^2 \left(\frac{T}{10^{9}\,{\rm K}}\right)^6 F \left(x_{\rm n}\right) \, , \\
Q_{\gamma^\prime}^{\rm pp} & = & \frac{328\pi^2}{93}\frac{31}{3780\pi} g^{\prime 2} \left( \frac{m_{\gamma'}^3}{T^2 \pi_L } \right)^2 T^8  \nn \\
&&\times  \left(\frac{f}{m_\pi}\right)^4 |\vec p_{F,{\rm p}}|^3 H\left(x_{\rm p}\right) \nn \\
& \simeq & 1.38\times 10^{12} \, {\rm erg}\,{\rm cm}^{-3}{\rm s}^{-1}  \left(\frac{|\vec p_{F,{\rm p}}|}{1/{\rm fm}}\right)^3 \nn \\
&&\times \left(\frac{1/{\rm fm}}{|\vec p_{F,e}|}\right)^{4} \left(\frac{10}{3 \ln\left(\min\left[\frac{|\vec p_{F,e}|}{m_e}, \frac{T}{m_{\gamma^\prime}}\right]\right)}\right)^2 \nn\\
&& \times \left(\frac{g^\prime m_{\gamma^\prime}}{10^{-9}\,{\rm MeV}}\right)^2 \left(\frac{T}{10^{9}\,{\rm K}}\right)^8 H \left(x_{\rm p}\right) \, ,
\eea
where $f\simeq 1$, $m_\pi$ is the pion mass, and
\bea
F\left(x\right) & = & 1- \frac{3x}{2}\arctan x^{-1} +\frac{x^2}{2(1+x^2)} \, ,\\
H\left(x\right) & = & \frac{2}{3} + \frac{5}{2}x^2 - \frac{x}{2}\left(3+5x^2\right)\arctan x^{-1} 
\eea
with $x_{\rm N} = m_\pi /2 |\vec p_{F,{\rm N}}|$.
Here, the neutron bremsstrahlung is only valid for the ${\rm U}(1)_{B-L}$ gauge boson scenario.

Now, one can explicitly show that  the emission of the dark gauge bosons from the PBF process are much stronger than that from the nucleon bremsstrahlung for $T\lesssim T_c$.  
The ratios of the emission rates are estimated as  
\bea \label{ratios}
&&\frac{Q_{\gamma'}^{\rm pp}}{Q_{\gamma^\prime, {\rm L}}^{{\rm p}^1\! S_0}}
\sim \Big(\frac{10}{v_{F,{\rm p}}^{4}}\Big) \Big(\frac{|\vec p_{F,{\rm p}}|}{m_\pi}\Big)^2 \Big(\frac{T}{m_{\rm p}}\Big)
\Big(\frac{T}{m_\pi}\Big)^2,\nonumber\\
&&\frac{Q_{\gamma'}^{\rm nn}}{Q_{B-L}^{{\rm n}^3\! P_2}}\sim 
\Big(\frac{1}{v_{F,{\rm n}}^{4}}\Big) \Big(\frac{m_{\rm n}}{m_\pi}\Big)\Big(\frac{T}{m_\pi}\Big)^3,
\eea
which are commonly of an order of $\mathcal{O}(10^{-7})$ at the critical temperatures, $T_c ({\rm n}^3\! P_2) = \mathcal{O}(0.1)\,{\rm MeV}$, so totally negligible. 

We note that the ratio of the neutrino emissivity from the nucleon bremsstrahlung to that from the PBF process is given by $Q_{\nu\bar{\nu}}^{\rm NN} / Q_{\nu\bar{\nu}}^{{\rm n}^3\! P_2} \sim 0.1 (m_{\rm N}/m_\pi)^3 (T/m_\pi) = \mathcal{O}(10^{-2})$.  It is also noticed that the neutrino emission of the modified Urca process is of an order of magnitude stronger than that of the nucleon bremsstrahlung. Therefore, at the critical temperature, the total bulk neutrino luminosity increases by a factor of ${\cal O}(10)$ due to the emergence of the PBF process. 
Comparing with this standard cooling scenario, the effect of the PBF process on the gauge boson emission is much more significant. 
Because of this, depending on the gauge boson mass,  Cas A can provide an even stronger constraint on the ${\rm U}(1)_{B-L}$ gauge coupling than  the low density stars (See Fig.~\ref{fig:B-LConstraints}).  
In the next section we discuss among others the possibility that such a dramatic amplification of the ${\rm U}(1)_{B-L}$ emission rate by the PBF process could help to fit the rapid cooling of Cas A appropriately.

\section{Analysis and Discussion}
\label{sec:AnD}

The observed effective surface temperature of Cas A  (including its rapid cooling rate)  and NS1987A is well fitted by the standard cooling scenario.  More specifically, to account for Cas A, there must be a superfluid phase transition in the core, in particular a weak neutron triplet superfluidity, and a lower amount of accreted light elements in the envelope~\cite{Wijngaarden:2019tht}. In the case of NS1987A, the superfluidity is less important, but a larger amount of light elements in the envelope is necessary for a better fit~\cite{Page:2020gsx}. 

If the emission rate of novel particles is close to or even greater than that of the neutrinos, a distinct deformation of the cooling curve occurs compared with the standard cooling history. This leads to constraints on the properties of the new particles. However, in the absence of a clear prospect to determine a particular set of fitting parameters,
it is also possible that the cooling curves can be successfully reproduced with the help of additional energy losses by new particles,  
even when the observation cannot be explained solely by the given NS model parameters within the standard cooling scenario.

For this reason, when such a modified cooling curve never fits the data in any choice of the NS model parameters, it gives an inevitable, i.e., {\it conservative} constraint on a novel particle. On the other hand, if a novel particle emission is effectively captured by the proper fitting parameters in the range that the observations are hardly explained by the standard cooling scenario, we will interpret it as a {\it hint} for a new particle beyond the SM. 
 
In the following subsections, we examine the simulation results for the cooling history of NS in the presence of the dark gauge bosons, and discuss their implications.

\subsection{Cooling simulation and input }

We utilize the public code `\texttt{NSCool}'\,\cite{nscool} for performing cooling simulations and modify it by adding the dark photon luminosity  obtained in the previous sections.
We employ the Akmal-Pandharipande-Ravenhall (APR)~\cite{Akmal:1998cf} equation of state (\texttt{`APR-EOS-Cat.dat'}) and assume a NS mass as $1.7 M_\odot$ \footnote{ This choice for the NS mass is a bit higher than the estimated mass range of $(1.22\text{--}1.62 )M_\odot$ for NS1987A. However, for low mass cases as $M_{\rm NS}< 1.9 M_\odot$ in the APR equation of state, the cooling curves do not significantly change because the fast cooling source by the direct Urca neutrino process is kinematically blocked for low temperatures.}.

Chemical compositions of the envelope can be characterized by $\eta \equiv g_{14}^2\Delta M/M$ \cite{Potekhin:1997mn}, where $g_{14}$ is the surface gravity in units of $10^{14} {\rm cm}/{\rm s}^{2}$ and $\Delta M$ is the accreted mass of light elements.
It turns out that $\eta$ value is one of the most important model parameters for the NS cooling simulation.
Since there is no clear criterion to determine the amount of light elements in the envelope for each NS,
we consider a broad range of $\eta$ from $10^{-13}$ as a thin layer of light elements, which is chosen in Ref.~\cite{Shternin:2010qi} to fit the Cas A data within the standard cooling scenario, to the very large value $10^{-3}$. Within the standard cooling scenario, the inferred thermal luminosity of NS1987A favors a rather thick layer of light elements\footnote{Because the explosion energy of SN1987A is expected to be substantially smaller than that of Cas A~\cite{Page:2020gsx}, it is natural that NS1987A has a thicker layer of light elements in the envelope.} like $\eta \geq 10^{-8}$.

Understanding the nuclear physics in extreme conditions is not precise enough yet so, in principle, there are various candidates for the nucleon gap profile.
In this paper, we select the specific model of the singlet pairing for the neutrons and the protons.
We pick the `\texttt{SFB}'~\cite{Schwenk:2002fq} model for the neutron singlet pairing.
But the choice of different models gives a tiny difference in the result because the pairing  occurs in the crust of NS, where the only dripped neutrons constitute a low density medium.
For the proton singlet pairing, we choose one of two models denoted as `\texttt{CCDK}'~\cite{Chen:1993bam} and `\texttt{T73}'~\cite{Taka:1993} which can be considered as the upper and lower limits of the possible model values for the gap, respectively.
The sensitivity of the simulations to the choice of the proton singlet pairing models is discussed at the end of  Sec.~\ref{sec:Results}.  

Meanwhile, the neutron triplet pairing plays an more important role to understand the rapid cooling of Cas A. 
It is noticed that the neutron triplet gap models in the literature are very broadly distributed~\cite{Ho:2014pta}. 
This implies the ${\rm n}^3\! P_2$ profile is highly uncertain, so it can be considered as a fitting parameter. 
In this paper, we take a phenomenological approach instead of taking a specific neutron triplet pairing model. 
The gap profile is approximated by a Gaussian with parameters determining its shape such as the height of the peak (equivalently, a critical temperature $T_c ({\rm n}^3\! P_2)$), width, and position in the momentum space as in Ref.~\cite{Hamaguchi:2018oqw}. 
This can be easily implemented in the public code `\texttt{NSCool}'~\cite{nscool}.  Among the parameters of the Gaussian gap profile, the gap height is most important to determine the onset time of the phase transition and the cooling rate;  the width and the position of the profile are still significant but less effective unless their values are taken to be extreme.

Our approach differs from Refs.~\cite{Ho:2014pta,Wijngaarden:2019tht}, where the NS mass (with the respective radius) is taken as 
the fitting parameter for each triplet pairing gap model.
However, these differences are effectively compensated by performing consistent
fitting of the Cas A data with appropriate Gaussian gap parameters of neutron triplet for a given NS mass.
 


\subsection{Results}\label{sec:Results}

\begin{figure}[t!]
\centering
  \includegraphics[width=1\linewidth]{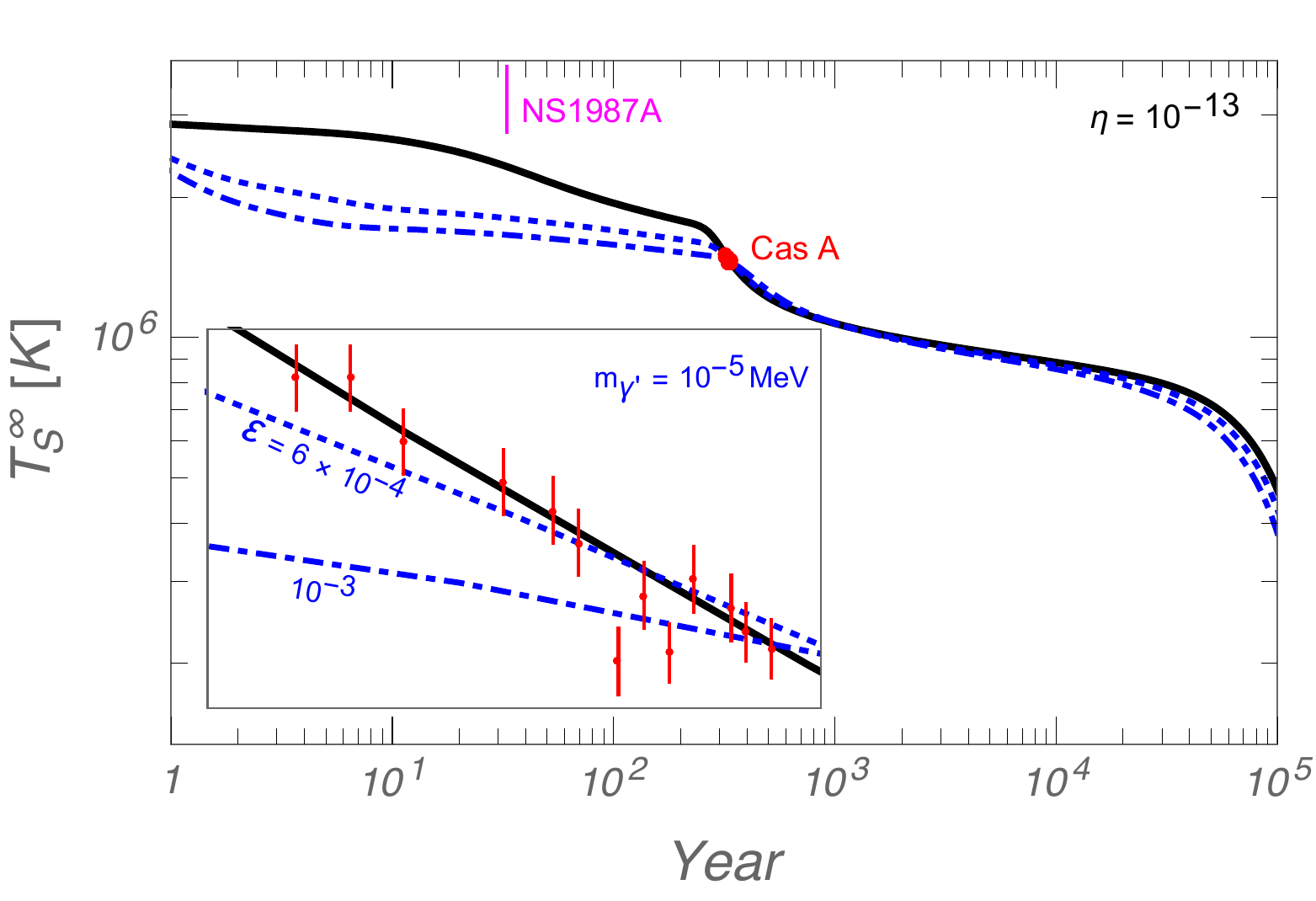}
  \includegraphics[width=1\linewidth]{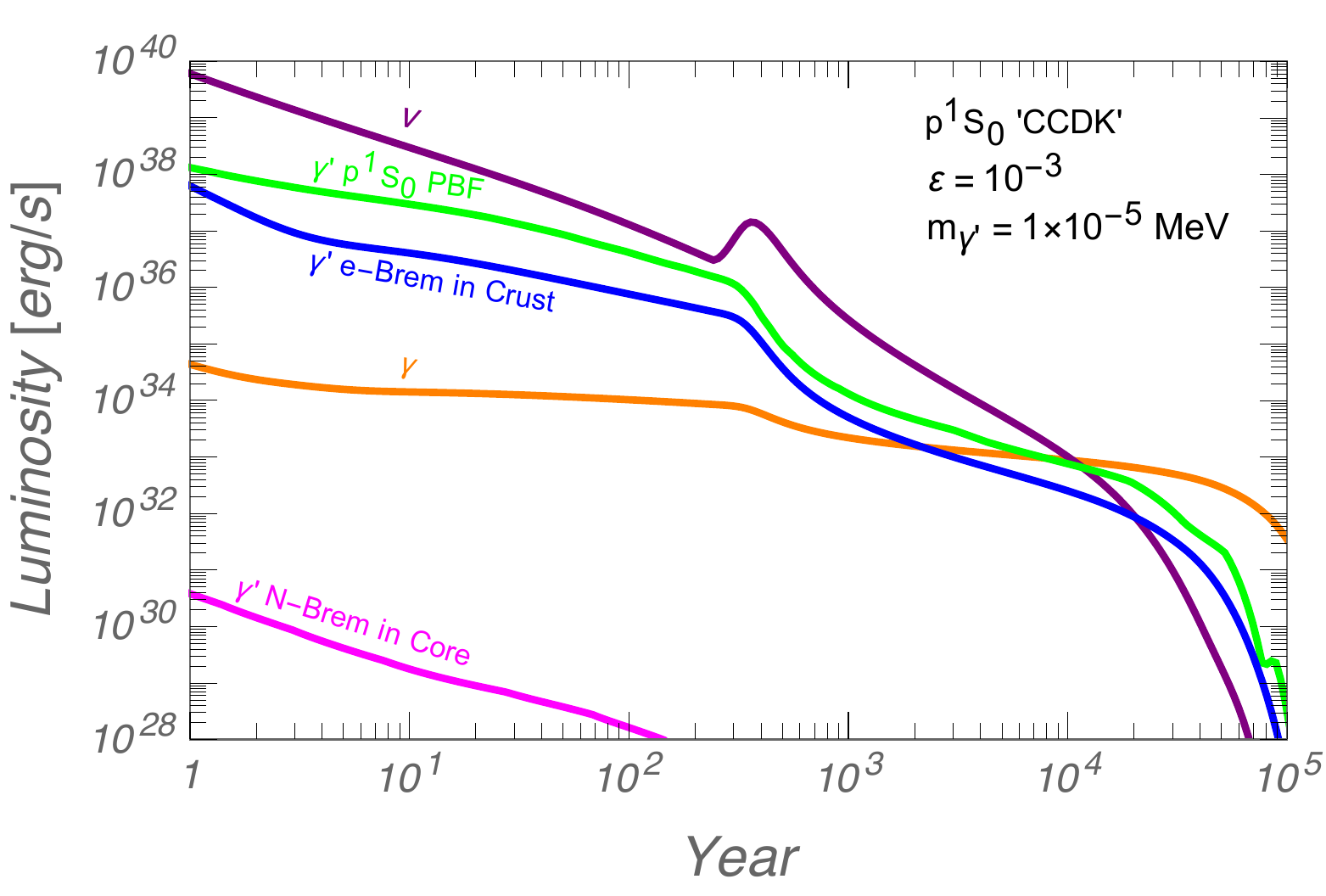}
\caption{\textit{Upper}: Cooling curves in the dark photon scenario for the parameter choice of $m_{\gamma^\prime} = 10^{-5}\,{\rm MeV}$ and $\varepsilon = 0$ (black), $6\times 10^{-4}$ (blue dashed), and $1\times 10^{-3}$ (blue dot-dashed) with the $\texttt{CCDK}$ model for the proton singlet pairing.
The red dots with the respective error bar indicate the redshifted surface temperatures implied by the Cas A data and the magenta dot corresponds to the inferred thermal temperature of the neutron star remnant of the SN1987A.
\textit{Lower}: Evolution of luminosity of the total neutrino emission (purple), the photon emission (orange), and the each dark photon emission processes with PBF (green), nucleon Bremsstrahlung in core (magenta), and electron Bremsstrahlung in crust (blue) for the parameter choice of $m_{\gamma^\prime} = 1\times 10^{-5}\,{\rm MeV}$ and $\varepsilon = 1.5\times 10^{-3}$.}
\label{fig:DarkPhotonAnalysis}
\end{figure}

\begin{figure}[t!]
\centering
  \includegraphics[width=1\linewidth]{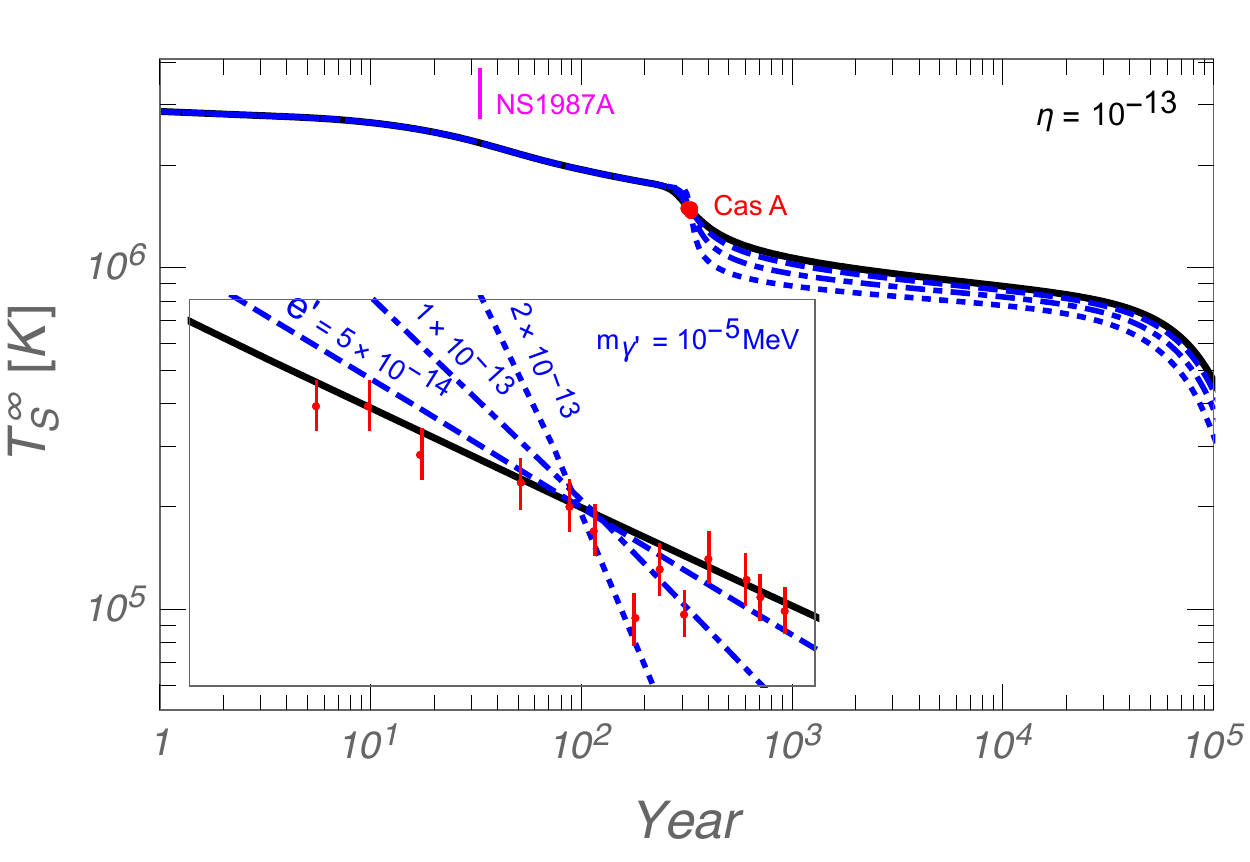}
  \includegraphics[width=1\linewidth]{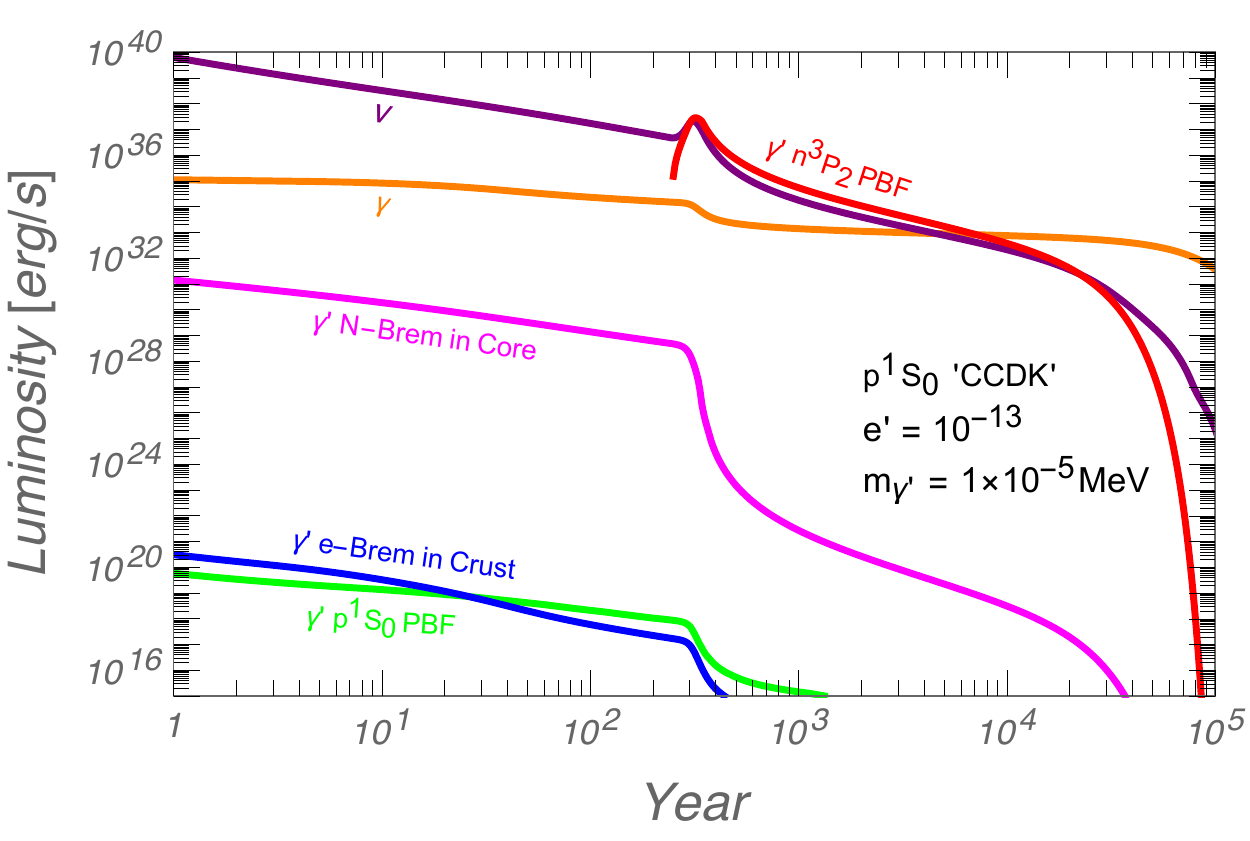}
\caption{\textit{Upper}: Cooling curves in the ${\rm U}(1)_{B-L}$ gauge boson scenario for the parameter choice of $m_{\gamma^\prime} = 10^{-5}\,{\rm MeV}$ and $e^\prime = 0$ (black), $5\times 10^{-14}$ (blue dashed), $1\times 10^{-13}$ (blue dot-dashed), and $2\times 10^{-13}$ (blue dotted) with the $\texttt{CCDK}$ model for the proton singlet pairing.
The red dots with the respective error bar indicate the redshifted temperature implied by the Cas A data and the magenta dot corresponds to the inferred thermal temperature of the neutron star remnant of the SN1987A.
\textit{Lower}: Evolution of luminosity of the total neutrino emission (purple), the photon emission (orange), and the each dark photon emission processes of ${\rm n}^3\!P_2$ PBF (red), ${\rm p}^1\!S_0$ PBF (green), nucleon Bremsstrahlung in core (magenta), and electron Bremsstrahlung in crust (blue) for the parameter choice of $m_{\gamma^\prime} = 1\times 10^{-5}\,{\rm MeV}$ and $e^\prime = 1\times 10^{-13}$.}
\label{fig:B-LAnalysis}
\end{figure}

The upper panels of Fig.~\ref{fig:DarkPhotonAnalysis} and Fig.~\ref{fig:B-LAnalysis} show the best fit curves (of the redshifted effective surface temperature $T_{\rm S}^\infty$) to the Cas A observations in the dark photon or ${\rm U}(1)_{B-L}$ gauge boson scenario with the specific proton singlet pairing profile denoted by `\texttt{CCDK}' and the assumption of the thin layer of light elements ($\eta = 10^{-13}$).
The black solid line corresponds to the case of the null hypothesis, i.e. the standard cooling scenario without any dark gauge boson emission~\cite{Page:2004fy,Yakovlev:2004iq,Page:2009fu}.
As we can see, it describes the Cas A data well.
The additional volume emission due to the dark gauge bosons therefore potentially changes the thermal history of NS.

Let us now discuss the constraints on the dark photon from the cooling of Cas A first. 
As shown in the upper panel of Fig.~\ref{fig:DarkPhotonAnalysis}, an energetic discharge of the dark photon could refrigerate NS efficiently that a cooling curve never traverses the observed points, making the data fitting futile.
Consequently, there is still a possibility to explain the Cas A data when
\bea
\varepsilon m_{\gamma^\prime} < 1.5\times 10^{-8}\,{\rm MeV}
\label{eq:DP_keyresult}
\eea
if $m_{\gamma^\prime} < T_c ({\rm n}^3\! P_2) = \mathcal{O}(0.1)\,{\rm MeV}$.
Here, the dark photon mass range of the constraint is determined by the critical temperature of the neutron triplet pairing because the critical temperature of the proton singlet pairing is typically larger.

The same simulation is done in the case of the ${\rm U}(1)_{B-L}$ gauge boson scenario.
The simulation result of the thermal evolution path is shown in the upper panel of Fig.~\ref{fig:B-LAnalysis}.
The constraint on the parameters of the ${\rm U}(1)_{B-L}$ gauge boson from the Cas A observation is given by
\bea
e^\prime < 1\times 10^{-13}
\label{eq:B-L_keyresult}
\eea
if $m_{\gamma^\prime} < T_c ({\rm n}^3\! P_2) = \mathcal{O}(0.1)\,{\rm MeV}$ and $\eta = 10^{-13}$.
Since the ${\rm U}(1)_{B-L}$ gauge boson couples to the neutron without the suppression by plasmon mass, the bound of Eq.~\eqref{eq:B-L_keyresult} has no mass dependence.
Indeed, the constraint on the dark photon of Eq.~\eqref{eq:DP_keyresult} moderates for a lower mass because the dark photon couples to the SM particles only through the kinetic mixing with the photon, which leads to the unavoidable plasma screening effect.

\begin{figure}[t!]
\centering
  \includegraphics[width=1\linewidth]{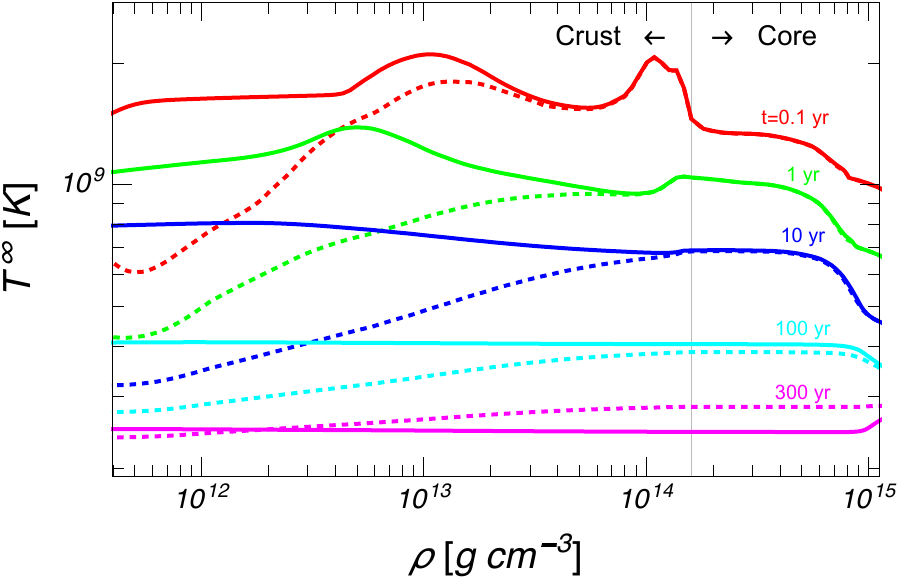}
\caption{Redshifted temperature profiles of the neutron star in the case of the null hypothesis (solid lines) and the dark photon scenario for $\varepsilon = 10^{-3}$ and $m_{\gamma^\prime} = 10^{-5}\,{\rm MeV}$ (dashed lines) in the respective age.}
\label{fig:ThermalRelaxation}
\end{figure}

The main yardstick of evaluating the constraint on the novel particle from Cas A  is the dominance of its luminosity against the neutrino.
The detailed luminosity history of each dark gauge  boson emission process in the respective scenario with the \texttt{CCDK} profile is exhibited in the lower panel of Fig.~\ref{fig:DarkPhotonAnalysis} (the dark photon case with $\varepsilon = 1.5\times 10^{-3}$ and $m_{\gamma^\prime} = 10^{-5}\,{\rm MeV}$) and Fig.~\ref{fig:B-LAnalysis} (the ${\rm U}(1)_{B-L}$ gauge boson case with $e^\prime = 1\times 10^{-13}$ and $m_{\gamma^\prime} = 10^{-5}\,{\rm MeV}$).
In the latter case, the luminosity of the dark gauge boson becomes comparable to the neutrino~\cite{Yakovlev:1998wr} at the age of Cas A, so the cooling curve descends manifestly as shown in the upper panel of Fig.~\ref{fig:B-LAnalysis}.
Before the neutron superfluid phase transition occurring at about 300\,yr, the volume emission rate of the ${\rm U}(1)_{B-L}$ gauge boson is negligible and therefore gives little effect on the early stage of thermal evolution.

In the case of the dark photon, the parametric choice of $\varepsilon = 1.5 \times 10^{-3}$ and $m_{\gamma^\prime} = 10^{-5}\,{\rm MeV}$ is constrained, although the corresponding volume emission rate is an order smaller than that of the neutrino at the age of Cas A.
This is mainly because the dark photon emission in the crust is relatively more significant compared with the ${\rm U}(1)_{B-L}$ gauge boson case.
At the stage of initial evolution up to 20 years, the core and the crust are not thermally equilibrated due to the low heat conductivity~\cite{Lattimer:1994glx}, hence the cooling of the core is not reflected in the surface temperature until the thermal relaxation is completed.
In fact, the surface temperature is mainly determined by the thermal properties in the crust.
Fig.~\ref{fig:ThermalRelaxation} shows the redshifted temperature profile in the respective age of the neutron star.
The solid lines and the dashed lines, respectively, correspond to the null hypothesis and the dark photon scenario with the parametric choice of $\varepsilon = 10^{-3}$ and $m_{\gamma^\prime} =10^{-5}\,{\rm MeV}$.
Compared to the result of the null hypothesis, Fig.~\ref{fig:ThermalRelaxation} clearly shows that  the crust of the neutron star at the early stage ($< 200$ years) could be chilled intensively by a sizable dark photon emission leading to a small but distinguishable deformation of the cooling curve at the age of Cas A ($\sim 300$ years).
Here, the core temperature at 300 years in the dark photon scenario is higher because the ${\rm n}^3\! P_2$ superfludity is weaker for the cooling curve to cross the Cas A data appropriately.

\begin{figure}[t!]
\centering
  \includegraphics[width=1\linewidth]{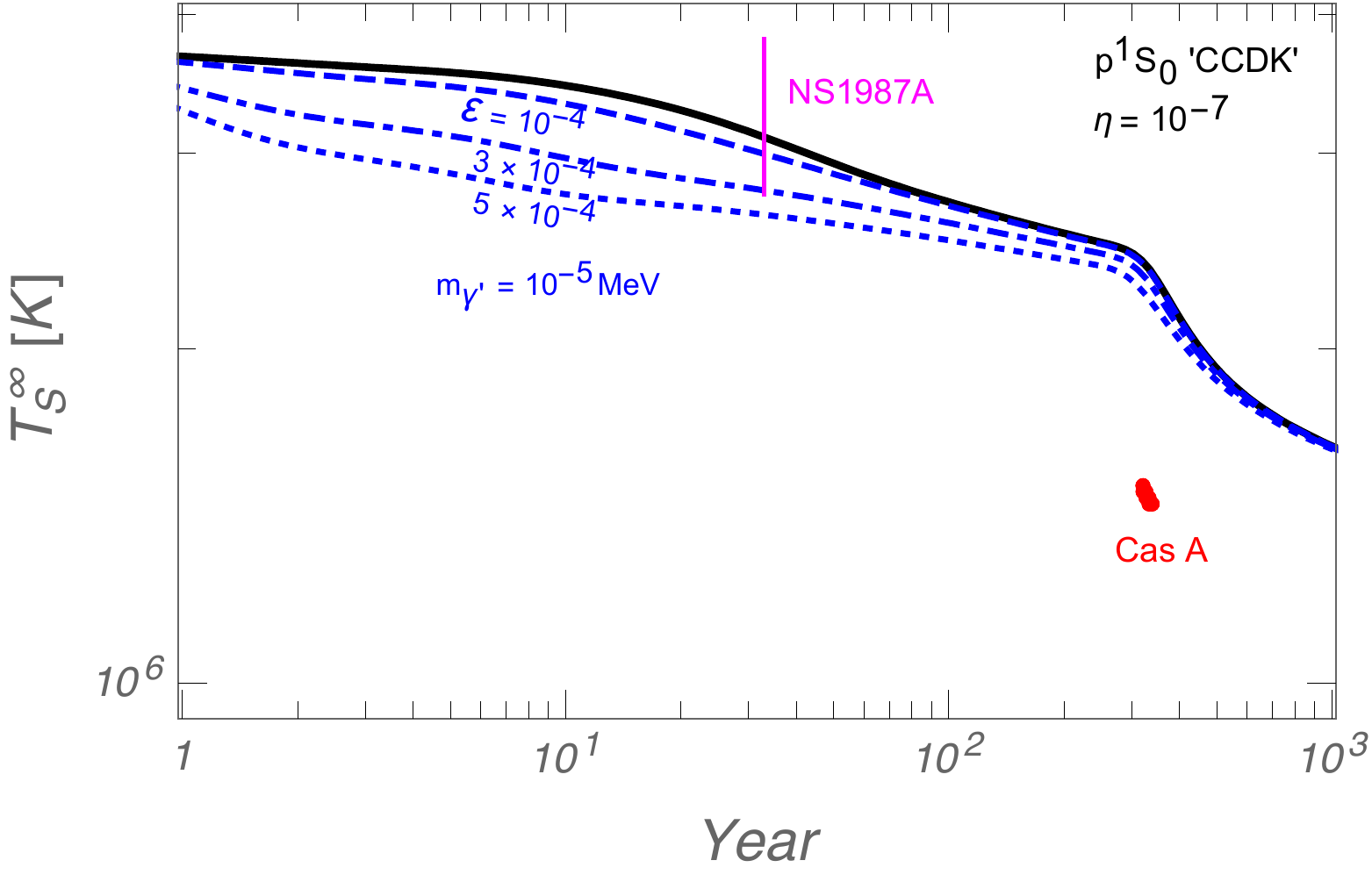}
\caption{Cooling curves in the dark photon scenario for the parameter choice of $m_{\gamma^\prime} = 10^{-5}\,{\rm MeV}$ and $\varepsilon = 0$ (black), $1 \times 10^{-4}$ (blue dashed), $3\times 10^{-4}$ (blue dot-dashed), and $5\times 10^{-4}$ (blue dotted) with the \texttt{CCDK} model for the proton singlet pairing and the rather thick layer of light elements given by $\eta = 10^{-7}$.
The red dots with the respective error bar indicate the redshifted surface temperatures implied by the Cas A data and the magenta line corresponds to the range of the inferred thermal temperature of the neutron star remnant of the SN1987A.}
\label{fig:DPforNS1987A}
\end{figure}

On one hand, 
in Fig.~\ref{fig:ThermalRelaxation}, one can see that the dark photon cooling effect becomes stronger at the lower density due to the smaller plasma suppression. Therefore, such a density-sensitive dark photon cooling in the crust could imprint a definite decrement on the cooling curves as shown in the upper panel of Fig.~\ref{fig:DarkPhotonAnalysis}.
In this sense, a very young neutron star like NS1987A ($t_{\rm NS} \sim 30$\,yr) is the good source to research the hidden signal of the dark photon.
In Fig.~\ref{fig:DPforNS1987A}, the magenta line marks the expected range of the redshifted effective surface temperature of NS1987A. It matches well to the standard cooling scenario with the rather thick layer of light elements ($\eta = 10^{-7}$) depicted by the black line.
If we believe the result of NS1987A, we can extract the constraint on the dark photon given by
\bea
\varepsilon m_{\gamma^\prime} < 3\times 10^{-9}\,{\rm MeV}
\label{eq:NS1987AResult}
\eea
for $m_{\gamma^\prime} \lsim \mathcal{O}(0.1)\,{\rm MeV}$ and this constraints is  one order more severe than the result from the Cas A data given by Eq.~\eqref{eq:DP_keyresult}.

\begin{figure}[t!]
\centering
  \includegraphics[width=1\linewidth]{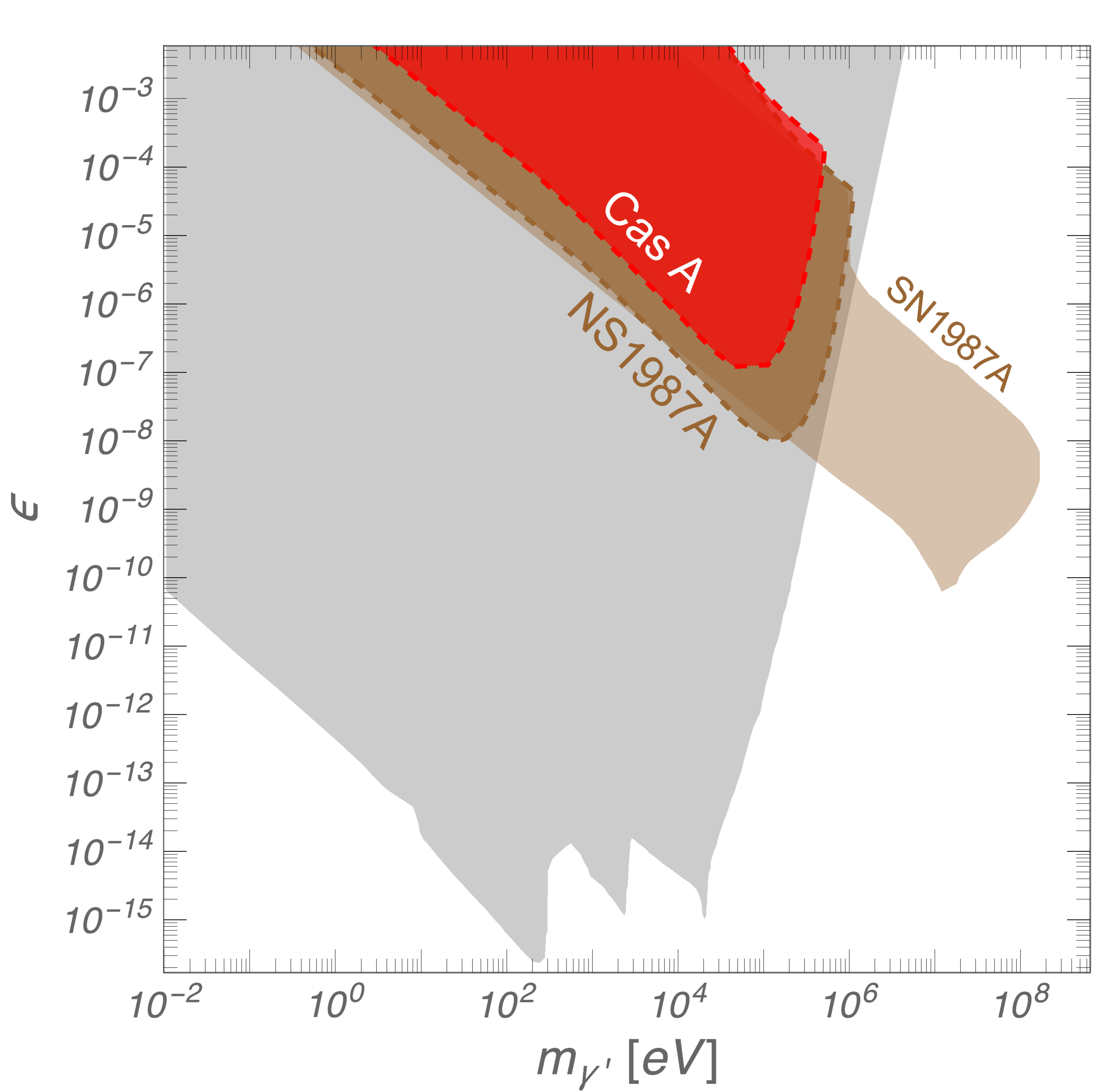}
\caption{The constraints on the dark photon scenario.
The red region is excluded by the rapid cooling of the neutron star in Cas A.
The darker brown region is excluded by the recent observation of the remnant of SN1987A.
The gray region indicates the preexisting constraints from stellar cooling argument in the sun, the HB stars, and the red giants~\cite{An:2013yfc,Hardy:2016kme,Redondo:2013lna,Fabbrichesi:2020wbt}.
The lighter brown region is from the stellar cooling argument in the first 10 seconds of SN1987A~\cite{Chang:2016ntp}.}
\label{fig:DPConstraints}
\end{figure}

\begin{figure}[t!]
\centering
  \includegraphics[width=1\linewidth]{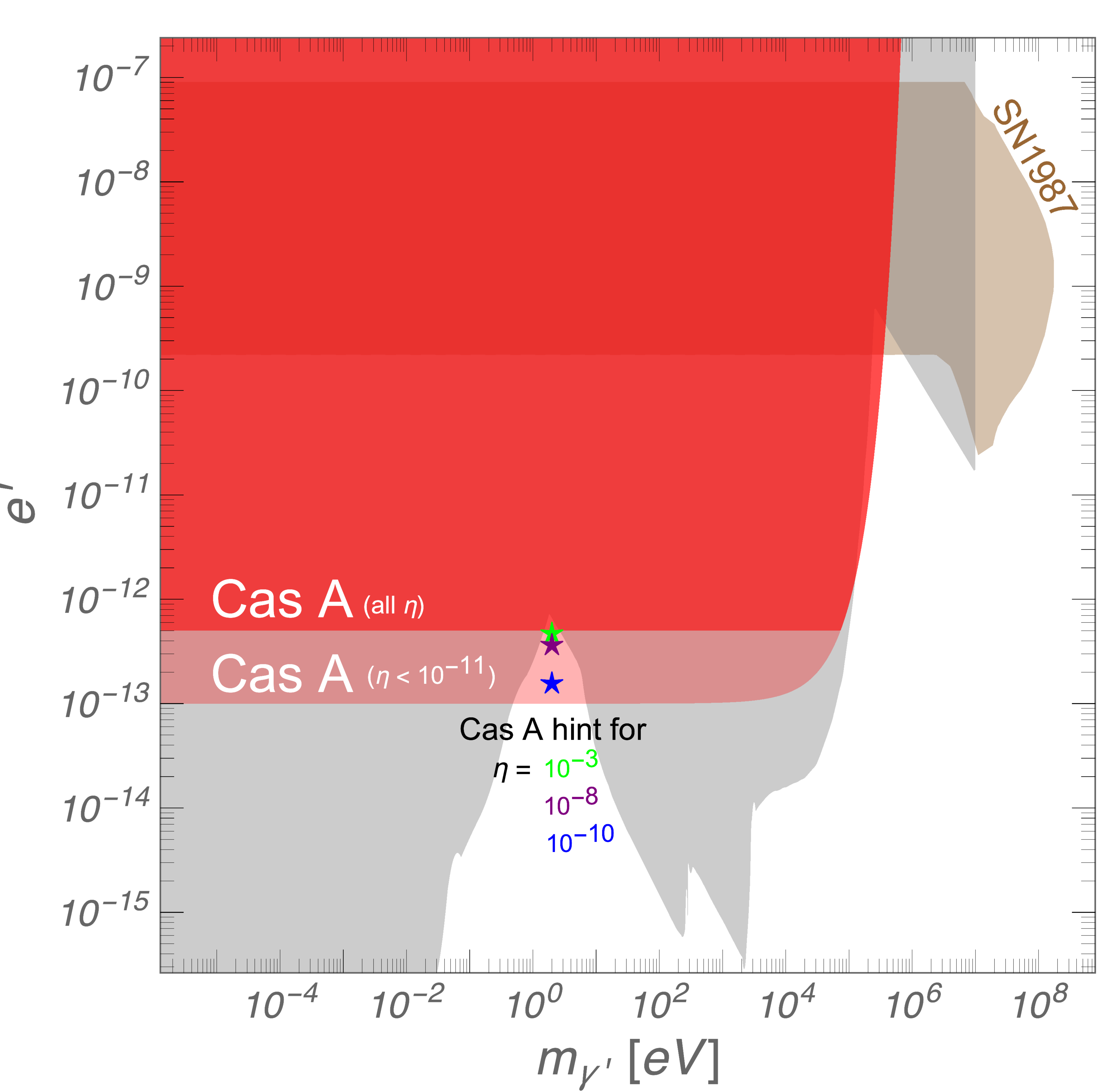}
\caption{The constraints plot for the ${\rm U}(1)_{B-L}$ gauge boson scenario.
The red region is conservatively (i.e. regardless of $\eta$) excluded by the rapid cooling of the neutron star in Cas A.
The Cas A bound can be extended up to the light red region when a lower $\eta$ ($< 10^{-11}$) is considered so the standard cooling scenario still gives an adequate fit to the data.
The gray region indicates the preexisting constraints from the fifth force searches~\cite{Murata:2014nra}, from BBN~\cite{Heeck:2014zfa,Knapen:2017xzo}, and from stellar cooling argument in the sun, the HB stars, and the red giants~\cite{Hardy:2016kme,An:2014twa}.
The lighter brown region is from the stellar cooling argument in the first 10 seconds of SN1987A~\cite{Knapen:2017xzo}.
The each colored star corresponds to the benchmark parametric choices in the respective $\eta$ values where the Cas A observation is described adequately  by the bulk ${\rm U}(1)_{B-L}$ gauge boson emission (shown in Fig.~\ref{fig:B-LforNS1987A}; see its text for details) without any conflict with the known observational constraints.}
\label{fig:B-LConstraints}
\end{figure}

Fig.~\ref{fig:DPConstraints} and Fig.~\ref{fig:B-LConstraints} show the constraints plot for the dark photon and the ${\rm U}(1)_{B-L}$ gauge boson scenario, respectively.
The plots contain the other astrophysical constraints (e.g. the stellar cooling argument for the sun, the red-giants, the horizontal branches, SN1987A and the fifth-force constraint) and also the cosmological bound such as the BBN bound.

If the dark gauge boson coupling  to the SM particles is large enough, then the energy transport via the dark gauge boson emission becomes inefficient due to trapping by the medium for the short mean free path compared to the geometric dimension of the star.
In the case of the dark photon scenario, the inverse ${\rm p}^1\!S_0$ PBF process ($A^\prime_\mu \rightarrow \tilde{\rm p}\tilde{\rm p}$) turns out to be the dominant absorption channel for the dark photons produced in the core\footnote{When $\varepsilon m_{\gamma^\prime} > 7 \times 10^{-6}\,{\rm MeV}$, the mean free path of the dark photon in the core becomes smaller than the radius of the neutron stars $\sim 10\,{\rm km}$.} but, nonetheless the emission rate in the crust is comparable and the produced dark photons can easily escape the crust because of its relatively short thickness. 
This  leads to somewhat weak lower limit on the dark photon coupling from the observations of Cas A and NS1987A. 
In the case of the ${\rm U}(1)_{B-L}$ gauge boson, if   $e'$ is bigger than $2.3 \times 10^{-9}$, the absorption by the inverse ${\rm n}^3\!P_2$ PBF process ($A^\prime_\mu \rightarrow \tilde{\rm n}\tilde{\rm n}$) becomes gradually important as the age of NS approaches that of Cas A.
However, if $e'$ is as large as $10^{-9}-10^{-3}$, the emissions through the neutron-bremsstrahlung process in the core or the electron scattering in the crust will distort the cooling curves before the neutron superfluid phase transition occurs. 
We estimate that the bound from the cooling of NS1987A becomes effective for $e'\gtrsim 10^{-7}$, and the ${\rm U}(1)_{B-L}$ gauge bosons are well trapped inside NS, so safe from the cooling constraints for $e^\prime \gtrsim 10^{-3}$.
For this reason, the upper limit from Cas A  and the bound from  NS1987A are not shown in Fig.~\ref{fig:B-LConstraints}.

\begin{figure}[t!]
\centering
  \includegraphics[width=1\linewidth]{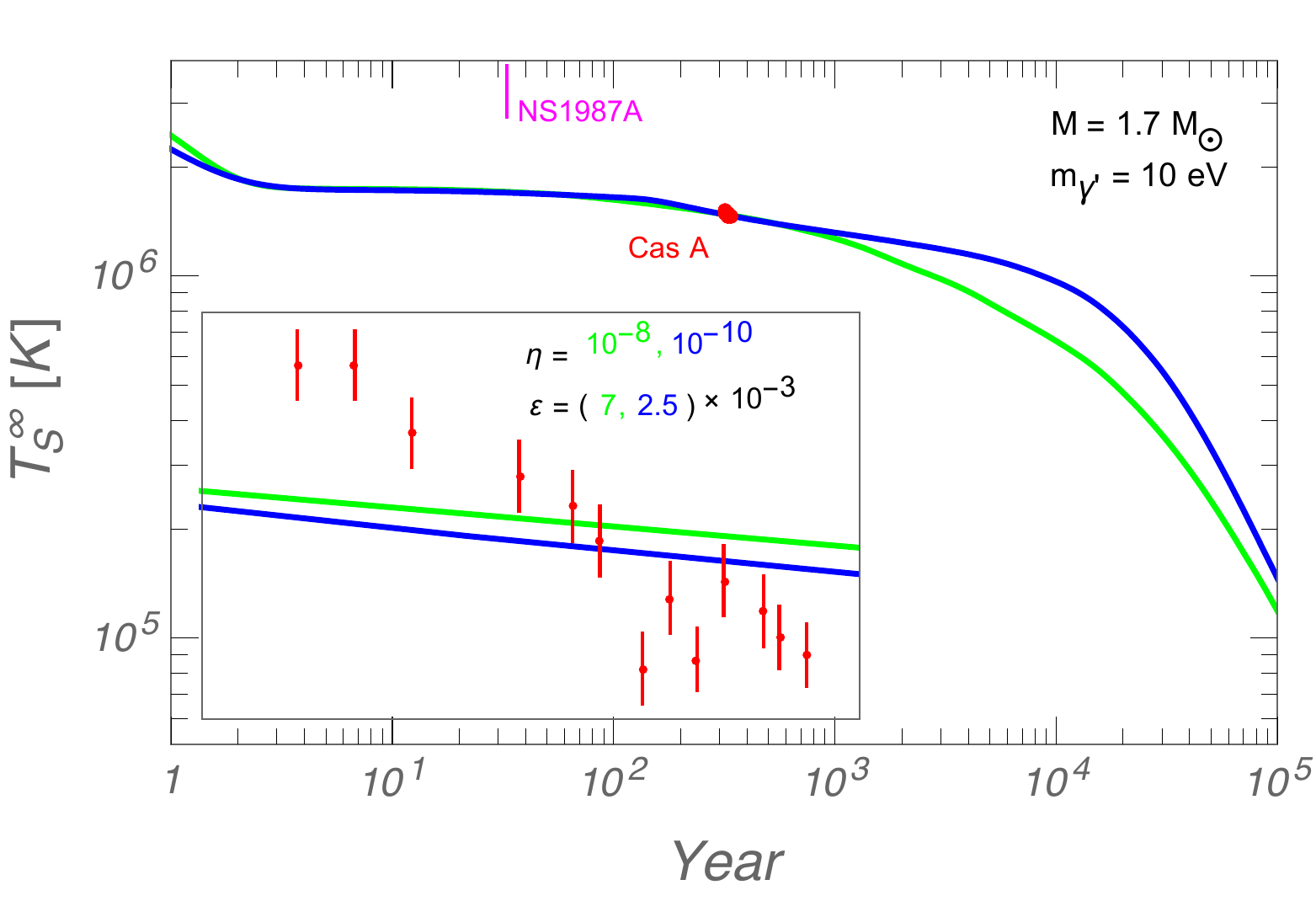}
  \includegraphics[width=1\linewidth]{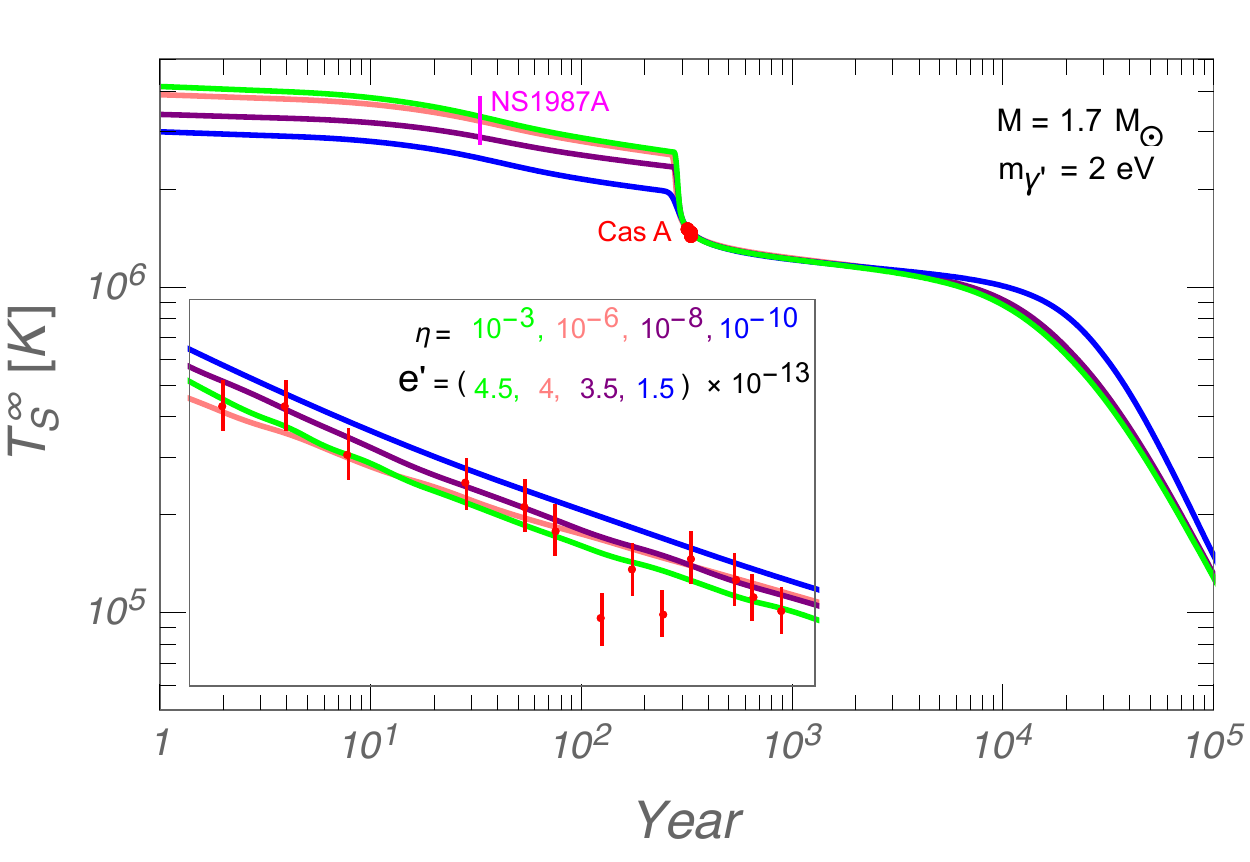}
\caption{ \textit{Upper}: Cooling curves in the dark photon scenario for the parameter choice of $m_{\gamma^\prime} = 10\,{\rm eV}$ and $e^\prime = 7\times 10^{-3}$ (green) and $2.5\times 10^{-3}$ (blue) with $\eta = 10^{-8}$ and $10^{-10}$, respectively, and the \texttt{CCDK} model for the proton singlet pairing.
The red dots with the respective error bar indicate the redshifted temperature implied by the Cas A data and the magenta line corresponds to the range of the inferred thermal temperature of the neutron star remnant of the SN1987A.
\textit{Lower}: Cooling curves in the ${\rm U}(1)_{B-L}$ gauge boson scenario for the parameter choice of $m_{\gamma^\prime} = 2\,{\rm eV}$ and $e^\prime = 4.5\times 10^{-13}$ (green), $4\times 10^{-13}$ (orange), $3.5\times 10^{-13}$ (purple), and $1.5\times 10^{-13}$ (blue) with $\eta = 10^{-3}$, $10^{-6}$, $10^{-8}$, and $10^{-10}$, respectively, and the \texttt{CCDK} model for the proton singlet pairing.}
\label{fig:B-LforNS1987A}
\end{figure}

\begin{figure}[t!]
\centering
  \includegraphics[width=1\linewidth]{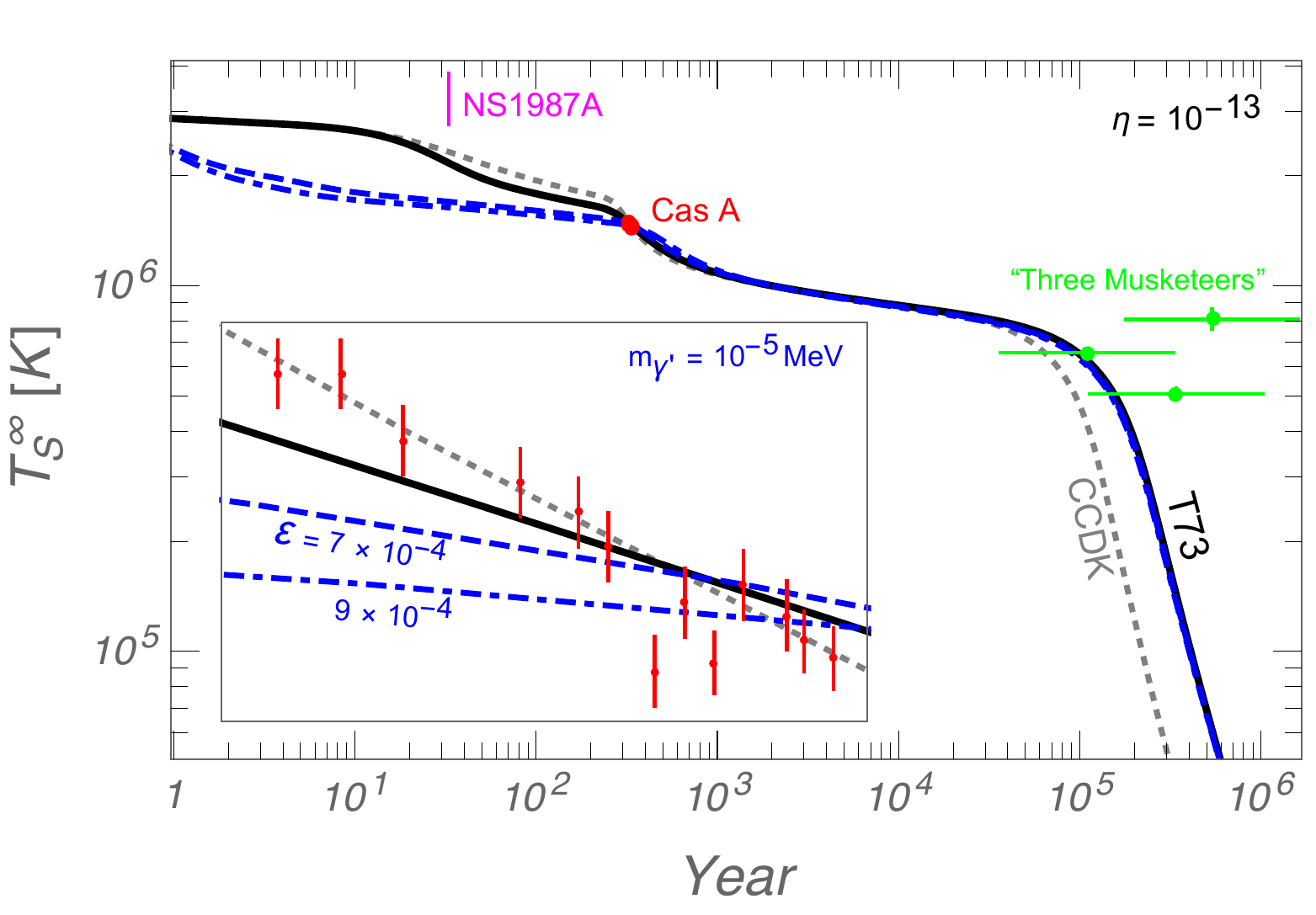} 
  \includegraphics[width=1\linewidth]{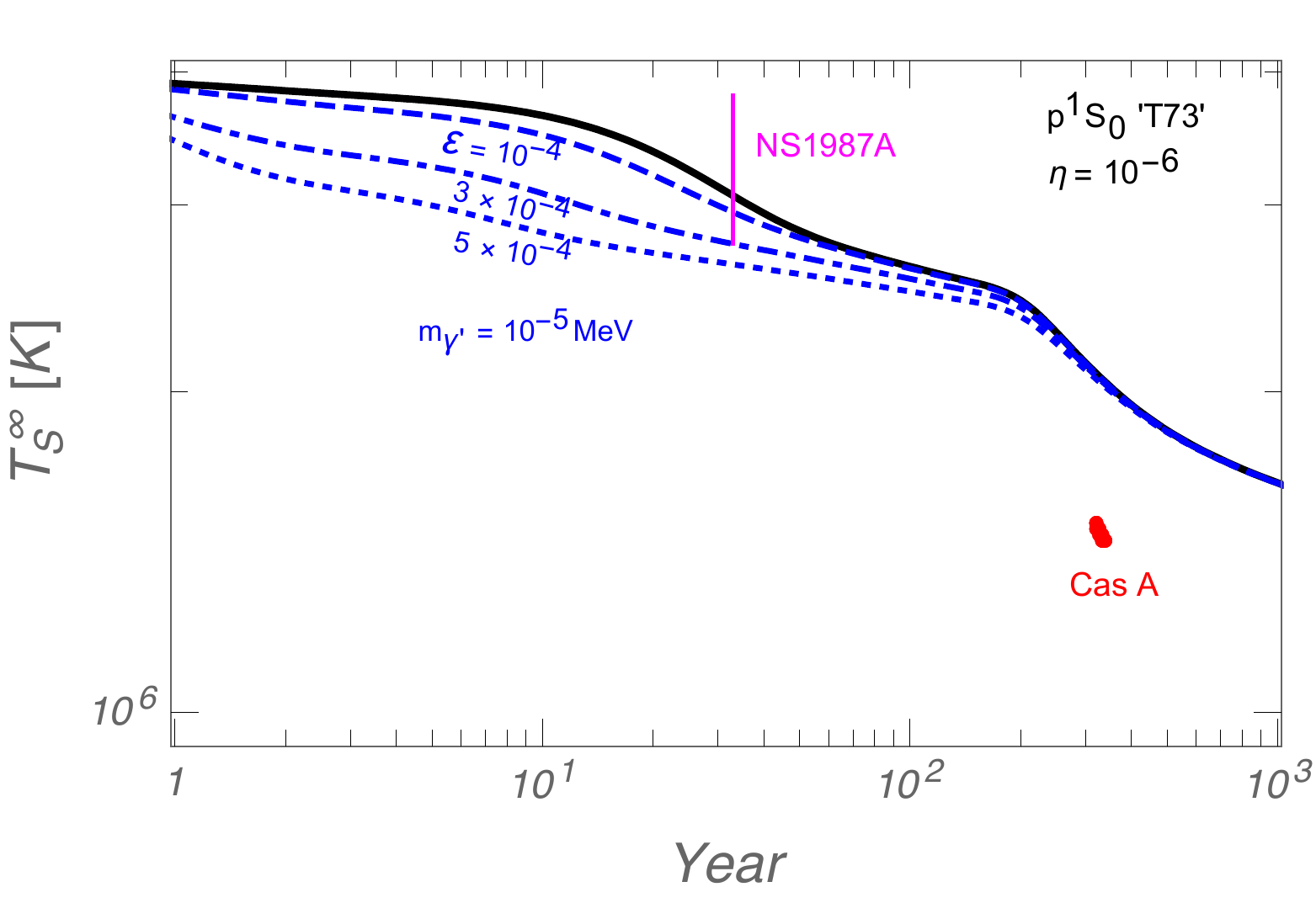}   
\caption{\textit{Upper}: Cooling curves in the dark photon scenario for the parameter choice of $m_{\gamma^\prime} = 10^{-5}\,{\rm MeV}$ and $\varepsilon = 0$ (black), $7\times 10^{-4}$ (blue dashed), $9\times 10^{-4}$ (blue dot-dashed) with the \texttt{T73} model for the proton singlet pairing and $\eta = 10^{-13}$.
The green dots with the error bar are for the `three Musketeers'~\cite{DeLuca:2004ck} of PSR B0656+14, Geminga, and PSR B1055-52.
The gray dotted line denoted by \texttt{CCDK} is the best fit curve for null hypothesis with the \texttt{CCDK} model to the Cas A observation.
\textit{Lower}: Cooling curves in the dark photon scenario for the parameter choice of $m_{\gamma^\prime} = 10^{-5}\,{\rm MeV}$ and $\varepsilon = 0$ (black), $1 \times 10^{-4}$ (blue dashed), $3\times 10^{-4}$ (blue dot-dashed), and $5\times 10^{-4}$ (blue dotted) with the \texttt{T73} model for the proton singlet pairing and $\eta = 10^{-6}$.}
\label{fig:T73coolingcurve}
\end{figure}

If the envelope contains a rather large amount of light elements, such as $\eta \geq 10^{-8}$ from the cooling implication of NS1987A, it leads to a higher temperature for the same interior temperature due to a higher thermal conductivity.
As shown in Fig.~\ref{fig:DPforNS1987A}, one can see that the black line corresponding to the standard cooling scenario with $\eta = 10^{-7}$ crosses the limit of NS1987A well, but a fitting to the observation of Cas A becomes much harder.
More specifically,  it is difficult to fit the Cas A data without new contributions for $\eta > 10^{-11}$~\cite{Ho:2014pta}.

Interestingly, in this case, the emission of the ${\rm U}(1)_{B-L}$ gauge bosons through the ${\rm n}^3\! P_2$ PBF process could support a fitting to the rapid cooling of Cas A. 
In the lower panel of Fig.~\ref{fig:B-LforNS1987A}, it is shown that the cooling curves of the ${\rm U}(1)_{B-L}$ gauge boson scenario with the proper choices of $e^\prime$ fit well the Cas A data  for a wide range of $\eta$, and even agree with NS1987A if $\eta > 10^{-8}$. 
This is due to the mild emission of ${\rm U}(1)_{B-L}$ gauge bosons from the crust, while the opportunely forceful emission occurs from the core at the age of Cas A. 

The colored star symbols (green, purple, blue) in Fig.~\ref{fig:B-LConstraints} denote these parametric choices of $e^\prime =$ ($4.5 \times 10^{-13}$, $3.5 \times 10^{-13}$, $1.5 \times 10^{-13}$) with $m_{\gamma^\prime} = 2\,{\rm eV}$ for the respective $\eta =$ ($10^{-3}$,  $10^{-8}$, $10^{-10}$) and they do not conflict with the known constraints on ${\rm U}(1)_{B-L}$ gauge bosons.
This may indicate important implications of the ${\rm U}(1)_{B-L}$ gauge boson.
As a consequence, the conservative constraint on ${\rm U}(1)_{B-L}$ gauge bosons from the rapid cooling observation of Cas A is given by
\bea
e^\prime < 5\times 10^{-13}
\label{eq:B-L_keyresult_conserv}
\eea
for $m_{\gamma^\prime} < T_c ({\rm n}^3\! P_2) = \mathcal{O}(0.1)\,{\rm MeV}$; the red region in Fig.~\ref{fig:B-LConstraints} is excluded conservatively.
The light red region in Fig.~\ref{fig:B-LConstraints} corresponding to
\bea
1\times 10^{-13} < e^\prime < 5\times 10^{-13}
\label{eq:B-L_keyresult_hint}
\eea
can be further excluded by Cas A if $\eta < 10^{-11}$, otherwise it implies the evidence of the ${\rm U}(1)_{B-L}$ gauge boson.

On the other hand, in the dark photon scenario, as manifested in the upper panel of Fig.~\ref{fig:B-LforNS1987A}, the dark photons cannot take on a role to provide a good fit to the data in the case of $\eta >  10^{-11}$ because their couplings to the SM particles in a medium are significantly suppressed by the plasma effect (for eletrons and protons) or even vanishes in the tree level (for neutrons). 
Since a rather small amount of light elements in the envelope ($\eta< 10^{-11}$) is required to optimise a fit to Cas A in the dark photon scenario, the given result of Eq.~\eqref{eq:DP_keyresult} could be considered as the conservative bound.

As the final remark, 
we also simulate cooling curves for the other proton singlet pairing model of \texttt{T73} and such results are presented at Fig.~\ref{fig:T73coolingcurve}.
In the upper panel, we choose the thin layer of light elements given by $\eta = 10^{-13}$.
Compared to the \texttt{CCDK} case (the gray dotted line), the fitting to the Cas A data (the black line) in the null hypothesis is less accomplished but better confident  to older and somewhat hot NS such as the three Musketeers~\cite{DeLuca:2004ck} (green).
There are two main reasons for this ${\rm p}^1\!S_0$ dependence: (i) variation of the heat capacity, and (ii) the gap profile ($\Delta$, a width) differences.
During the superfluid and superconducting phase transitions, the heat capacity of medium jumps up discontinuously at $T=T_c$ (so called `Lambda point') followed by an exponential reduction at lower $T$ due to a phase space suppression of excited states.
Before the thermal relaxation, the cooling curves are manifestly identical irrespective of the gap profiles for the ${\rm p}^1\!S_0$ pairing.
Once the interior of NS gets thermally relaxed by energy transfer from the crust to the core, the cooling history is mainly determined by the core, and the model dependence  is revealed. 
The \texttt{CCDK} profile consists of the shape with a larger gap and a wider width than that of \texttt{T73}, so discontinuity of the heat capacity is more significant. 
Accordingly, the heat capacity for the \texttt{CCDK} profile is slightly larger at the early stage, but it gets smaller subsequently for $t_{\rm NS} \gtrsim 10^4$\,yr.
Therefore, the cooling curve of the \texttt{CCDK} profile is above that of the \texttt{T73}  
 for $t_{\rm NS}\sim 10-100$\,yr, while it is located below for $t_{\rm NS} >10^4$\,yr as shown in Fig.~\ref{fig:T73coolingcurve}.
In the \texttt{T73} case, we estimate that the constraints on the dark photon from the three Musketeers is given as $\varepsilon m_{\gamma^\prime} < 5\times 10^{-8}\,{\rm MeV}$ for $m_{\gamma^\prime} < 10^{-3}\,{\rm MeV}$ which is a little bit milder than the constraints from the Cas A data in the \texttt{CCDK} case given in Eq.~\eqref{eq:DP_keyresult}.

In the lower panel of Fig.~\ref{fig:T73coolingcurve}, the rather large amount of light elements in the envelope ($\eta=10^{-6}$) is considered to fit the observation of NS1987A.
As discussed in the \texttt{CCDK} case, the dark photon emission in the crust through the electron bremsstrahlung may give a significant effect on the early thermal history of NS before the thermal relaxation so that we can get a stringent bound on the dark photon from NS1987A.
Since the only neutron singlet pairing affects to the cooling of the crust and we assume the same neutron singlet pairing profile (\texttt{SFB}), we get the similar result to the \texttt{CCDK} case given by Eq~\eqref{eq:NS1987AResult} which could mean its robustness.

\section{Conclusions}
\label{sec:con}

We have examined the young NS cooling observations to figure out constraints or hints on new light gauge bosons.
In order to avoid physical uncertainties such as age or contamination from heating processes,
we pick the two specific young NS,  Cas A and NS1987A to perform cooling simulations, rather than including a comprehensive list of shinning NS.

As with studies on other astrophysical objects, the NS cooling simulations suffer several theoretical uncertainties.
For instance, we have assumed the specific NS mass as $1.7M_{\odot}$.
This value is within the range ($< 1.9\, M_{\odot}$) where no dramatic change happens in the cooling curves for different values of NS mass. Therefore, it leads to a consistent fitting to the data, and we confirmed it.
The other important uncertainty arises from the amount of accreted light elements in the envelope (parametrized by $\eta$), which affects the relation between the internal and the surface temperature.
 In principle, a wide range of $\eta$ values can be considered because of no clear observable evidence to determine $\eta$.
While an envelope with a thin layer of light elements fits the Cas A data properly within the standard cooling scenario~\cite{Shternin:2010qi}, the envelope of NS1987A is expected to have a rather thick layer due to its lesser supernova explosion energy and somewhat larger inferred X-ray luminosity.


There are additional uncertainties associated with the microscopic theories in a dense medium; the neutron superfluidity and the proton superconductivity.
The neutron singlet-state superfluidity in the crust and the proton singlet-state superconductivity in the core give a relatively small effect on the NS thermal history (and consequently on our results).
More specifically, for the dark photon scenario, its production is contributed not only by the proton singlet PBF in the core, but also by the electron bremsstrahlung in the crust which relies on the electron coupling. For the ${\rm U}(1)_{ B-L}$ gauge boson scenario, its emission is dominated by the neutron triplet PBF.
The neutron triplet pairing gap profile has a large uncertainty so that we consider it as a complementary fitting parameter to the NS mass. This approach is easily accessible in NS cooling simulations.
We adopt a Gaussian gap profile in the Fermi momentum space, with a height $\sim 5 \times 10^{8}\,{\rm K}$ (for the right onset time to match the age of Cas A) and a wide width to cover all the core.

We find that there are the two dominant dark gauge boson production processes: the nucleon PBF in the core and the electron bremsstrahlung with scattering off heavy nuclei in the crust.
If $\varepsilon m_{\gamma^\prime} < 1.5 \times 10^{-8}\,{\rm MeV}$ for the dark photon scenario and $e^\prime < 5 \times 10^{-13}$ for the ${\rm U}(1)_{B-L}$ gauge boson scenario with $m_{\gamma^\prime} < T_c({\rm n}^3\! P_2) = \mathcal{O}(0.1)\,{\rm MeV}$, the volume emission of dark gauge bosons carries out little alteration of the cooling curve in the standard cooling scenario which fits well to the Cas A data.
The existence of NS1987A as a compact remnant of SN1987A gives the robust constraint on the dark photon physics; when $\varepsilon m_{\gamma^\prime} < 3\times 10^{-9}\,{\rm MeV}$ for   $m_{\gamma^\prime}  \lesssim 0.1-1\,{\rm MeV}$, NS1987A is well described by the speculated thermal luminosity.
Together with the trapping condition for a large coupling regime, Fig.'s ~\ref{fig:DPConstraints} and \ref{fig:B-LConstraints} show the resulting bound on the dark photon scenario and the ${\rm U}(1)_{B-L}$ scenario, respectively, from NS1987A and the rapid cooling of Cas A.

Recent temporal observation of Cas A might also provide a hint on the ${\rm U}(1)_{B-L}$ gauge boson.
As already discussed in several literatures,
a thin layer of light elements in the envelope should be chosen  in order for the standard cooling scenario to fit the Cas A data. This is because a lower heat transfer efficiency is more suitable for Cas A.
However, even if a larger amount of light elements were accreted in the envelope, a ${\rm U}(1)_{B-L}$ gauge boson emission through the ${\rm n}^3\! P_2$ PBF process could compensate for this effect and helps to fit the rapid cooling of Cas A due to its neutron-philic nature as shown in  the lower panel of Fig.~\ref{fig:B-LforNS1987A}. We find therefore that the parametric region of $1\times 10^{-13} < e^\prime < 5\times 10^{-13}$ can be further excluded by Cas A if $\eta < 10^{-11}$.  Turning this argument around, if $\eta$ is larger or $10^{-11}<\eta<10^{-3}$, we may argue that the rapid cooling of Cas A could imply the existence of the ${\rm U}(1)_{B-L}$ gauge boson with the mass around ${\cal O}({\rm eV})$, as shown in Fig.~\ref{fig:B-LConstraints}.

Further observations of the young NS will improve our understanding of NS, allowing us to examine microscopic theories in extreme circumstances. If future data consistently confirms the rapid cooling of Cas A, the bounds on dark gauge bosons obtained in this paper will become more conservative, or possibly indicate the presence of new particle emission (in particular, of the ${\rm U}(1)_{B-L}$ gauge boson). A more accurate observation of NS1987A will be of great interest as it directly detects the thermal relaxation phase in young NS. This can provide a robust constraint on the dark photon scenario.

\begin{acknowledgments}
We are grateful to Koichi Hamaguchi, Natsumi Nagata, Jiaming Zheng and Dany Page for the useful communications. 
This research was supported by Basic Science Research Program through the National Research Foundation of Korea (NRF) funded by the Ministry of Education (NRF-2017R1D1A1B06033701) (DKH) and also by IBS under the project code, IBS-R018-D1 (CSS).
The work of S.Y. is supported by the research grant ``The Dark Universe: A Synergic Multi-messenger Approach'' number 2017X7X85K under the program PRIN 2017 funded by the Ministero dell’Istruzione Università e della Ricerca (MIUR).
\end{acknowledgments}

\appendix



\end{document}